\documentclass[11pt,a4paper]{article}
\usepackage{jcappub}
\usepackage{lmodern}
\usepackage{graphicx,color}
\graphicspath{{figures/}}
\usepackage{comment}
\usepackage{cancel}

\usepackage[T1]{fontenc}
\usepackage[utf8]{inputenc}
\usepackage{color}
\usepackage{amsmath}
\usepackage{amssymb}
\usepackage{esint}
\usepackage{floatpag}
\usepackage{orcidlink}

\makeatletter

\definecolor{somegreen}{cmyk}{0,0.49,0.98,0.09}
\definecolor{red}{rgb}{1,0,0}
\definecolor{magenta}{cmyk}{0,1,0,0}
\definecolor{lavender}{cmyk}{0.16,0.67,0,0.57}
\definecolor{darkgreen}{rgb}{0,0.65,0.05}
\definecolor{antiquefuchsia}{rgb}{0.33, 0.1, 0.89}

\let\jnl@style=\rm
\def\ref@jnl#1{{\jnl@style#1}}

\def\aj{\ref@jnl{AJ}}                   
\def\actaa{\ref@jnl{Acta Astron.}}      
\def\araa{\ref@jnl{ARA\&A}}             
\def\apj{\ref@jnl{ApJ}}                 
\def\apjl{\ref@jnl{ApJ}}                
\def\apjs{\ref@jnl{ApJS}}               
\def\ao{\ref@jnl{Appl.~Opt.}}           
\def\apss{\ref@jnl{Ap\&SS}}             
\def\aap{\ref@jnl{A\&A}}                
\def\aapr{\ref@jnl{A\&A~Rev.}}          
\def\aaps{\ref@jnl{A\&AS}}              
\def\azh{\ref@jnl{AZh}}                 
\def\baas{\ref@jnl{BAAS}}               
\def\bac{\ref@jnl{Bull. astr. Inst. Czechosl.}}
\def\caa{\ref@jnl{Chinese Astron. Astrophys.}}
\def\cjaa{\ref@jnl{Chinese J. Astron. Astrophys.}}
\def\icarus{\ref@jnl{Icarus}}           
\def\jcap{\ref@jnl{J. Cosmology Astropart. Phys.}}
\def\jrasc{\ref@jnl{JRASC}}             
\def\memras{\ref@jnl{MmRAS}}            
\def\mnras{\ref@jnl{MNRAS}}             
\def\na{\ref@jnl{New A}}                
\def\nar{\ref@jnl{New A Rev.}}          
\def\pra{\ref@jnl{Phys.~Rev.~A}}        
\def\prb{\ref@jnl{Phys.~Rev.~B}}        
\def\prc{\ref@jnl{Phys.~Rev.~C}}        
\def\prd{\ref@jnl{Phys.~Rev.~D}}        
\def\pre{\ref@jnl{Phys.~Rev.~E}}        
\def\prl{\ref@jnl{Phys.~Rev.~Lett.}}    
\def\pasa{\ref@jnl{PASA}}               
\def\pasp{\ref@jnl{PASP}}               
\def\pasj{\ref@jnl{PASJ}}               
\def\rmxaa{\ref@jnl{Rev. Mexicana Astron. Astrofis.}}%
\def\qjras{\ref@jnl{QJRAS}}             
\def\skytel{\ref@jnl{S\&T}}             
\def\solphys{\ref@jnl{Sol.~Phys.}}      
\def\sovast{\ref@jnl{Soviet~Ast.}}      
\def\ssr{\ref@jnl{Space~Sci.~Rev.}}     
\def\zap{\ref@jnl{ZAp}}                 
\def\nat{\ref@jnl{Nature}}              
\def\iaucirc{\ref@jnl{IAU~Circ.}}       
\def\aplett{\ref@jnl{Astrophys.~Lett.}} 
\def\apspr{\ref@jnl{Astrophys.~Space~Phys.~Res.}}
\def\bain{\ref@jnl{Bull.~Astron.~Inst.~Netherlands}}
\def\fcp{\ref@jnl{Fund.~Cosmic~Phys.}}  
\def\gca{\ref@jnl{Geochim.~Cosmochim.~Acta}}   
\def\grl{\ref@jnl{Geophys.~Res.~Lett.}} 
\def\jcp{\ref@jnl{J.~Chem.~Phys.}}      
\def\jgr{\ref@jnl{J.~Geophys.~Res.}}    
\def\jqsrt{\ref@jnl{J.~Quant.~Spec.~Radiat.~Transf.}}
\def\memsai{\ref@jnl{Mem.~Soc.~Astron.~Italiana}}
\def\nphysa{\ref@jnl{Nucl.~Phys.~A}}   
\def\physrep{\ref@jnl{Phys.~Rep.}}   
\def\physscr{\ref@jnl{Phys.~Scr}}   
\def\planss{\ref@jnl{Planet.~Space~Sci.}}   
\def\procspie{\ref@jnl{Proc.~SPIE}}   

\makeatother

\title{Model independent approach towards measuring expansion and growth factor from next generation galaxy clustering and lensing angular power spectrum
}
\author[a,b,c]{Ziad Sakr \orcidlink{0000-0002-4823-3757} }
\affiliation[a]{Instituto de Física Teórica UAM-CSIC, Campus de Cantoblanco, 28049 Madrid, Spain}
\affiliation[b]{Institute of Theoretical Physics, Philosophenweg 16, Heidelberg University, 69120, Heidelberg, Germany}
\affiliation[c]{Faculty of Sciences, Universit\'e St Joseph; Beirut, Lebanon}
\date{}

\emailAdd{ziad.sakr@net.usj.edu.lb}

\abstract{In this work we perform Fisher forecasts on the expansion and the growth factors following model independent approaches from 3$\times$2pt joint analysis of the galaxy lensing, clustering, and their cross-correlated spectra at the linear, and extending as well to non linear scales. For that, instead of choosing a specific model for the matter power spectrum, the main ingredient of these probes, we express it  by parametrizing its components, such as the expansion and the growth factor, and those of the standard halo model and excursion set theory in several $z$ bins, besides to the different bias and non-linear bias modelling functions.
 We apply the technique to Euclid, Rubin and SKA public specifications in the range $0.2\le z \le 1.8$ and show that one can then obtain model-independent constraints of the expansion $E(z_i)$ and the growth factor $G(z_i)$. We also show the change in gain in precision at each $z$-shell when going from pessimistic cut at linear scales to more optimistic non-linear settings, or the difference between using each survey alone or a combination of all of them, or the impact from fixing or adding more degrees of freedom in the non-linear modeling. We found that, in the most agnostic case, one can still reach high precision on $E(z_i)$ in the order of the percent level when combining the three surveys at once while the growth factor $G(z_i)$ has for the same settings one order of magnitude weaker constraints. We also found for both factors, an improvement that can reach  one order of magnitude in precision when passing from linear to non-linear scales. We conclude that we will be able to constrain the two important factors of the background evolution and structure formation of the Universe when using non linear scales and the combined power of future surveys even in the most agnostic approaches.}

\begin{document}
\maketitle

\section{Introduction}

Large-scale galaxy surveys have always served in providing a better understanding of the physics governing our Universe. In the near future, both ground surveys like DESI~\citep{DESI:2016fyo}, 4MOST~\citep{4MOST2019}, J-PAS~\citep{Bonoli:2020ciz}, the Rubin Observatory Legacy Survey of Space and Time (LSST)~\citep{LSSTScience:2009jmu} and space surveys such as Euclid~\citep{laureijs2011euclid,Amendola:2016saw} and the Nancy Roman Space Telescope survey (Roman)~\citep{Eifler:2020hoy,Rose:2021nzt} will further expand our spatial and temporal knowledge by an order of magnitude, yielding sub-percent estimates of the main cosmological parameters. Together with other sources of information, from Hubble diagrams to CMB, the possibility of finding  deviations from the standard cosmological model, or its definitive confirmation, seems within reach.

One of the main difficulties in the data analysis of 3$\times$2pt galaxy clustering and lensing probes is the optimal exploiting of the non-linear information for cosmological purposes. Scales below some tens of Megaparsecs are  subject to non-linear evolution, but an accurate modelling of this evolution is hard to produce. In the last decades valiant efforts were put forward in developing both semi-analytical methods such as React, and effective field theory treatment of the effects of the short scales and large ones such as PPF \cite{Baumann:2010tm,Pietroni:2011iz,Carrasco:2012cv,Manzotti:2014loa},  and cosmological simulations, which often rely only on gravitational effects (for a recent review see \cite{Angulo:2021kes}) but more recently also on more sophisticated hydrodynamical implementations (see e.g.~\citep{Khandai:2014gta,Crain:2015poa,McCarthy:2016mry,Dave:2019yyq,Castro:2020yes,Barreira:2021ukk}) and fitting halo model to these simu. Nevertheless, these methods have been in fact developed and tested only for standard cosmologies and a few other selected cases. A restriction to a limited set of cosmologies could lead to missing new physics or to biasing the parameter estimation when analysing real data.

In this paper we perform forecasts for future surveys following a complementary route. Instead of focusing on specific cosmological models, we develop a methodology to extract from the data information that is as much model-independent as possible, following the approach that is partly and sparsely presented at the linear level in \cite{2020JCAP...11..054B}. More specifically, we do not assume a cosmological model neither at the background nor at the perturbed level, and parametrize galaxy/matter biasing on general grounds, via the parameterised bias expansion. We use a Fisher matrix analysis extended, in which the parameters are not the usual cosmological ones, but rather the linear growth factor $G(z)$ of the linear power spectrum and the background expansion factor $E(z)$ in redshift bins, plus free functions of linear and non linear halo bias, the halo model principal ingredients, i.e. the ones related to the spherical collapse or the halo mass function modelling. It is important to note that, as we shall see later, a matter density component $\Omega_{\rm m,0}$ is added to our list of free parameters, as being part and specific to the one halo term when modeling the non linear or the galaxy intrinsic alignment effect, with the latter being relevant for the lensing component of our used 3x2pt probe. All these elements combine into our theory-informed parametrization for the non-linear scales we apply to surveys that approximate the expected specifications of the Euclid, Rubin and SKA galaxy surveys,\footnote{Although for brevity we refer often to Euclid, Rubin and SKA surveys, in both cases it is understood that we are not providing official specifications but just use publicly available information.} covering the redshift range from 0.2 to 1.8.

Recently, other more or less similar approaches have been considered in the literature, with different degrees of model-independency, or usage of the clustering and lensing of the angular projection of large scale structures distribution at different redshifts, or extensions to non linear scales,  e.g. \cite{Srinivasan:2024nkv,Broxterman:2024oay,Ye:2024rzp,Sarmiento:2025yyh}.
The investigation of this paper, though more general than each of the above, is still preliminary because the theoretical combination of excursion set theory and halo model and nonlinear bias that we employ has been so far only tested in a limited number of cases beyond $\Lambda$CDM. It might well be therefore that our parametrization, even if largely more general than those based on specific cosmological models, turns out to be still insufficient to reproduce with high fidelity the non-linear behaviour. The method we describe here, however, can be directly improved to more general forms that might be developed in the future for instance. Therefore, we believe that, notwithstanding its current limitations, our method is a useful step forward.

The structure of the paper is the following: in Sec.~\ref{NLmodel} we lay down the modelling we followed in computing the theoretical predictions. Then in Sec.~\ref{probes} we describe the method and datasets we use to forecast constraints on our model independent parameters. Then in Sec.~\ref{results} we show the results for different cases and combinations of datasets before we discuss and conclude in Sec.~\ref{conclusion}.

\section{Fiducial Power spectrum at nonlinear scales}\label{NLmodel}

We will adopt a model where we split the two-point statistics into a 1-halo term (both points are located in the same halo) and a 2-halo term (the two points are located in different haloes) such as  $P_{\delta_{\rm m}\delta_{\rm m}}^{NL}(k,z) =P_{\delta_{\rm m}\delta_{\rm m}}^{\mathrm{1h}}(k,z)+P_{\delta_{\rm m}\delta_{\rm m}}^{\mathrm{2h}}(k,z)$.

\noindent The 1-halo term is:
\begin{equation}
P_{\delta_{\rm m}\delta_{\rm m}}^{\mathrm{1h}}(k,z) =  \int_{0}^{\infty} \mathcal{H}_{\mathrm{\delta}}^{2}(k, M, z) \, n(M,z) \, \mathrm{d} M\,,
\label{eqn:P1hss}
\end{equation}
where 
\begin{equation}
\mathcal{H}_{\mathrm{\delta}}(k, M, z) = {M \over \bar{\rho}_{\mathrm{m}}} \,  \tilde{u}_{\mathrm{h}}(k \vert M, z)\,,
\end{equation}
with $\bar{\rho}_{\rm m}$ being the present day mean matter density of the Universe and we assume that dark matter haloes are spherically symmetric on average, and have density profiles, $\rho(r \vert M, z) = M \, u_{\mathrm{h}}(r \vert M, z)$, that depend only on their mass $M$, and $u_{\mathrm{h}}(r \vert M)$ is the normalised density profile of a dark matter halo, with $\tilde{u}_{\mathrm{h}}(k \vert M)$ the Fourier transforms of the  halo density profile, both normalised to unity [$\tilde{u}(k \! \!= \!\! 0 \vert M) \!\! = \!\! 1 $]. We assume that the density profile of dark matter haloes follows a Navarro-Frenk-White (NFW) profile \citep{Navarro:1996gj}, described by two parameters the concentration, $c$, and mass, $M$, of the halo, however these two parameters are correlated. In this work we adopt for our fiducial the \cite{Duffy:2008pz} concentration-mass relation,
\begin{equation}
    \label{eq:con_duffy}
    c(M, z) = 10.14\; \ \left[\frac{M}{(2\times 10^{12} M_{\odot}/h)}\right]^{- 0.081}\ (1+z)^{-1.01} \ ,
\end{equation}
\\
 and $n(M)$ is the halo mass function in the following form: 
\begin{equation}\label{eq:hmf}
n(M,z) = \frac{\overline{\rho}_{\mathrm{m}}}{M^{2}} \nu f(\nu) \frac{\mathrm{d} \ln \nu}{\mathrm{d} \ln M}\,,
\end{equation}
with $\nu = \delta_{\mathrm{c}}(z) / \sigma(M, z)$ being the peak height. Here $\delta_{\mathrm{c}}$ is the critical overdensity required for spherical collapse at redshift $z$, and $\sigma(M, z)$ is the mass variance, while  $f(\nu)$ is the multiplicity function we elaborate on next.    

To determine $f(\nu)$ we adopt the excursion set approach that introduces a mass-dependent barrier of the form
\begin{align}
\label{eq:drifting_barrier}
B = \delta_c(z) + \beta S \: ,
\end{align}
where $S = \sigma^2(R(M),z)$.
In addition to the barrier drift, the collapse dynamics themselves are complicated by environmental effects and fuzzy halo definitions. In \cite{MaggioreII} this was taken into account by turning the barrier itself into a Gaussian stochastic variable with a mean $\bar B=\delta_c(z) + \beta S$ and width $D_B$. Both the trajectories and the barrier itself perform a random walk, and the  joint probability distribution is obtained from a 2D Fokker-Planck equation 
\citep{Corasaniti:2011dr}
\begin{equation}
\frac{\partial\, \Pi}{\partial S} = \frac{1}{2} \frac{\partial^2 \Pi}{\partial \, \delta^2} + \frac{D_B}{2} \frac{\partial^2 \Pi}{\partial B^2} \:,
\end{equation}
which is solved for $f (\sigma) = 2\sigma^2 \frac{\delta F(\sigma^2)}{\delta \sigma^2} $ where $F(\sigma^2) = 1 - \int_\infty^{\delta_c} \Pi(\delta,\sigma^2)d \delta$ to give

\begin{align}
\label{eq:multiplicity_general_barrier}
f_k(\sigma) = \sqrt{\frac{2a}{\pi}} \frac{1}{\sigma} \mathrm{e}^{-a \bar B^2/(2 \sigma^2)} \left( \bar B - \sigma^2 \frac{\mathrm{d} \bar B}{\mathrm{d} \sigma^2} \right) \: ,
\end{align}
where $a \equiv 1/(1+D_B)$. Using Eq.~\ref{eq:drifting_barrier}, this reduces to a Press-Schechter like solution when the constant threshold $\delta_c$ is replaced by the full barrier:
\begin{equation}
\label{eq:f_k_GR}
f_k(\sigma) = \sqrt{\frac{2 a}{\pi}} \frac{\delta_c}{\sigma} \mathrm{e}^{-a(\delta_c + \beta \sigma^2)^2 / 2 \sigma^2} \: .
\end{equation}

\noindent To account for realistic filter functions which leads to deviations from the simple assumed uncorrelated random walk, we need to introduce the real-space top-hat multiplicity function $f_x$, to first order in $\kappa$ \citep{Maggiore:2009rv,Kopp:2013lea},
\begin{align}
\label{eq:f_x_GR}
f_x(\sigma) = f_k(\sigma) + f_{1,\beta=0}^{m-m}(\sigma) + f_{\beta^{(1)}}^{m-m}(\sigma) + f_{1,\beta^{(2)}}^{m-m}(\sigma)
\end{align}
where the Markovian term $f_k$ for a diffusive, drifting barrier is given for the fiducial value by Eq.~\ref{eq:f_k_GR}, while the corrections are:
\begin{align}
f_{1,\beta=0}^{m-m}(\sigma) &= a \kappa \frac{\delta_c}{\sigma} \left( \mathrm{e}^{-a \delta_c^2 / 2 \sigma^2} - \frac{1}{2} \Gamma \Big( 0, \frac{a \delta_c^2}{2 \sigma^2} \Big) \right) \: ,\\
f_{\beta^{(1)}}^{m-m}(\sigma) &= -a \delta_c \beta \left(a \kappa \: \mathrm{erfc} \Big( \delta_c \sqrt{\frac{a}{2 \sigma^2}} \Big) + f_{1, \beta = 0}^{m-m} (\sigma) \right) \: , \\
f_{1,\beta^{(2)}}^{m-m}(\sigma) &= -a \beta \left( \frac{\beta}{2} \sigma^2 f_{1,\beta=0}^{m-m}(\sigma) + \delta_c f_{1,\beta^{(1)}}^{m-m} (\sigma) \right) \: ,\end{align}

\noindent with the parameterisation in the fiducial limit $\delta_c$ being well approximated by (Nakamura1997)
\begin{equation}
\label{eq:delta_c_{GR}}
\delta_c^\mathrm{fid}(z) = \frac{3 (12 \pi)^{2/3}}{20} \left(1 - 0.0123 \log_{10} \bigg( 1 + \frac{\Omega_m ^ {-1} - 1}{(1 + z)^3} \bigg) \right) \: .
\end{equation}
As noted in \cite{Hagstotz:2018onp}, all expressions are defined for the smoothed density field $\sigma^\mathrm{fid}$ calculated in a standard cosmology, the threshold is imposed on the initial conditions, and all subsequent effects of modified model are encapsulated in the dynamics of the barrier and its elements such as $\delta_c$, and the Markovian multiplicity function $f_k$.

\noindent The 2-halo terms are given, for a given redshift, by:
\begin{align}\label{P2h}
P_{\delta_{\rm m}\delta_{\rm m}}^{\mathrm{2h}}(k, z) = P_{\delta_{\rm m}\delta_{\rm m}}^{\mathrm{lin}}(k, z) \,  &\int_{0}^{\infty} \mathrm{d} M_1 \, \mathcal{H}_{\mathrm{\delta}}(k, M_1, z) \, b_{\mathrm{h}}(M_1, z)\, n(M_1) \nonumber \\\ 
&\times \int_{0}^{\infty} \mathrm{d} M_2 \, \mathcal{H}_{\mathrm{\delta}}(k, M_2, z) \, b_{\mathrm{h}}(M_2, z)\, n(M_2, z) \nonumber \\\
&+ P_{\delta_{\rm m}\delta_{\rm m}}^{\mathrm{lin}}(k, z) \, I_{\mathrm{NL}}(k, z) \,
\end{align}
where $P_{\delta_{\rm m}\delta_{\rm m}}^{\mathrm{lin}}(k)$ is the linear power spectrum, evaluated for the cosmological parameters at the fiducial: $\Omega_{\rm cdm}=0.270$, $\Omega_{\rm b}=0.049$,  $\Omega_k=0$, $h=0.67$, $n_s=0.96$, $\sigma_8=0.83$, and $b_\mathrm{h}(M,z)$ is the halo bias function given in our excursion set framework in the peak split model by 
\begin{equation}
\label{eq:linear_bias}
b_h(M,z) = 1 + \left( \frac{a \bar B}{S} - \frac{1}{\bar B - S \frac{\mathrm d \bar B}{\mathrm d S}} \right) \: ,
\end{equation}
while the second term in equation \ref{P2h} encompasses the beyond-linear halo bias correction $b_{\mathrm{NL}}$ proposed by (Mead2021) where, 
\begin{equation}
\begin{split}
I_{\mathrm{NL}}(k, z) & = \int_{0}^{\infty} \int_{0}^{\infty} \mathrm{d}M_1 \mathrm{d}M_2 \ b_{\mathrm{NL}}(k,M_1,M_2,z) \\
& \times \mathcal{H}_{\mathrm{\delta}}(k, M_1, z)\,\mathcal{H}_{\mathrm{\delta}}(k,M_2, z)\\
& \times n(M_1)\, n(M_2, z)\, b_{\mathrm{h}}(M_1, z)\, b_{\mathrm{h}}(M_2, z) \ .
\end{split}
\end{equation}
where $b_{\mathrm{NL}}$ is measured at the fiducial using the \textsc{DarkQuest} emulator \citep{Nishimichi:2018etk,Mahony:2022emy}, by measuring the non-linear halo-halo power spectrum and then dividing it by the linear matter power spectrum multiplied with the product of linear bias factors as in \cite{Mead:2020qdk}. We limit the wavenumber to $k\sim 7$ below the value for which the final ${P_k^{\rm NL}}$ was shown in \cite{Mahony:2022emy} to be in agreement with the one extracted from the simulations.

\section{Probes and specifications used in the Fisher matrix}\label{probes}

Given the relatively large redshift uncertainties that we expect from photometric measurements (compared to spectroscopic observations), these analyses will be performed via a tomographic approach, in which galaxies are binned into redshift slices that are considered as two-dimensional (projected) data sets.

\noindent The observed  angular lensing-lensing convergence
power spectrum from a survey divided into several redshift bins  can be written as

\begin{equation}
    C_{ij}^{\rm \gamma\gamma}(\ell)=c \int{\rm d}z\,\frac{W_i^{\rm \gamma}(z)W_j^{\rm \gamma}(z)}{H(z)r^2(z)}P_{\delta_{\rm m}\delta_{\rm m}}^{\rm NL}{\rm }(k,z)\,\label{eq:Cl_GCph}
\end{equation}
where $P_{\delta_{\rm m}\delta_{\rm m}}^{\rm NL}(k,z)$ is the  power spectrum, evaluated as in Sect.~\ref{NLmodel} at $k=k_{\ell}(z)=\frac{\ell + 1/2}{r(z)}$, \\
and
\begin{align}
W_i^{\rm \gamma}(k,z) =&\; \frac{3}{2}\Omega_{\rm m,0} \left(\frac{H_0}{c}\right)^2(1+z)r(z) \,
 \int_z^{z_{\rm max}}{dz'\frac{n_i(z)}{\bar{n}_i}\frac{r(z')-r(z)}{r(z')}} \, , \label{eq:wl_mg}
\end{align}
where $n_i(z)/\bar{n}_i$ and $b_i(k,z)$ are, respectively, the normalised galaxy distribution and the galaxy bias in the $i$-th redshift bin. For this evaluation we will assume
that the fiducial bias in $\Lambda$CDM is scale independent and equal
to $\sqrt{1+z}$.\\

\noindent We as well tried to include  intrinsic alignment effects within our formalism where the weight function can be written as
\begin{equation}
W_i^{\rm IA}(z) =  \frac{n_{i}(z)}{c/H(z)} = - \left( \frac{H_0}{c} \right) n_i(z) E(z){\cal{A}}_{\rm IA} {\cal{C}}_{\rm IA} \Omega_{{\rm m},0}\frac{{\cal{F}}_{\rm IA}(z)}{G(z,k)} ,
\label {eq: iaweight}
\end{equation}
with ${\cal{F}}_{\rm IA}(z) = (1 + z)^{\eta_{\rm IA}}[\langle L \rangle(z)/L_{\star} (z)]^{\beta_{\rm IA}}, $ where $\langle L_g(z)\rangle$ and $\langle L_{g\star}(z)\rangle$ are the redshift-dependent mean as computed from the luminosity function and the characteristic luminosity of source galaxies, respectively, $\eta_{IA}$ and $\beta_{IA}$ are the redshift and power law dependence parameters of the luminosity function while $\mathcal{A}_{IA}$ and $\mathcal{C}_{IA}$ are further nuisance parameters.\\
We then use $W_i^{\rm IA}(z)$ to replace $W_i^{\rm \gamma}(k,z)$ in Equ.~\ref{eq:Cl_GCph} in order to obtain either the correlation between background shear and foreground intrinsic alignment $C^{\rm I\gamma}_{ij}(\ell)$, or the autocorrelation of the foreground intrinsic alignment $C^{\rm II}_{ij}(\ell)$. \\

\noindent We also include the photometrically detected galaxy-galaxy correlations for which we define the radial weight function for galaxy clustering as
\begin{equation}\label{eq:photoGCwin}
W^{\rm G}_i (k,z)=b_i(k,z) n_i (z)\frac{H(z)}{c} \, ,
\end{equation}.\\
and use to replace $W_i^{\rm \gamma}(k,z)$ in order to obtain this time the galaxy-galaxy autocorrelation or the galaxy-galaxy lensing cross correlations.\\

\noindent The parameterizations division of our model independent parameters follows the redshift values of the 10 equipopulated binning  
\begin{equation}
z_i = \{0.418, 0.560, 0.678, 0.789, 0.900, 1.019, 1.155, 1.324, 1.576\},
\end{equation}
such that $p_{\alpha}^{\text{\tiny{GC}}}=\{G(z_{i}),E(z_{i}),\delta_c(z_{i}),\Delta_v(z_{i}),\beta(z_{i}),D_B(z_{i}),b(z_{i}),b_{NL}(z_{i},M_i), C_{NL}(z_{i},M_i), \dots\}$,
subscripts run over the $z$ bins. Additionally for
$b_{\rm NL}$ and $C$ in the case where we do not limit their variation to be only redshift dependent but also mass dependent, the latter binning would be the following
\begin{equation}
\log_{10} (M_i)=\{13.0,13.25,13.5,14.0,14.25,14.75,15.0\}.   
\end{equation}

The Fisher matrix for survey of photometric galaxy-galaxy clustering, galaxy-galaxy lensing, and their cross-correlation, that covers a fraction of the sky $f_{{\rm sky}}$, is a sum over $\ell$ bins \cite[see e.g.][]{Euclid:2019clj}
\begin{equation}
F_{\alpha\beta}^{\text{XC}}=\frac{1}{2}\sum_{\ell = \ell_{min}}^{\ell_{max}}(2\ell+1)\sum_{ABCD}\sum_{ij,mn}\frac{ C^{AB}_{ij}}{\partial p_{\alpha}}\left[\Delta C^{-1}(\ell)\right]^{BC}_{jm}\frac{ C^{CD}_{mn}}{\partial p_{\beta}}\left[\Delta C^{-1}(\ell)\right]^{DA}_{ni}\,,\label{eq:wlfm}
\end{equation}
where the block descriptors $A,B,C,D$ run
over the combined probes lensing and clustering and the indices $i,j,m,n$ are implicitly summed over, while 
\begin{align}
             \Delta{C}^{AB}_{ij}(\ell) &= \frac{1}{\sqrt{f_{\rm sky} \Delta \ell}}\left[C^{AB}_{ij}(\ell) + N^{AB}_{ij}(\ell)\right],
             \label{eq: covsecond}
       \end{align}  
where the parameters are $p_{\alpha}=\{G(z_i),E(z_i),\dots\}$.
Here, $\ell$ is being summed from $\ell_{{\rm min}} = 10.0$ to $\ell_{{\rm max}}(z)= k_{\rm max}  r(z) -1/2$, where $k_{\rm max} = 0.1$  or 0.2 or 0.5 or 0.7 $h \rm {Mpc}^{-1}$ since we considered different cases for different $k_{max}$ limits, with $\Delta\log\ell=0.1$.\\

To explore the dependence on the specifications, we also experiment with different datasets for our photometric analysis for which we consider :\\

\begin{itemize}
\item Euclid is a European Space Agency medium-class space mission due for launch in 2023. It will carry on-board a near-infrared spectrophotometric instrument and a visible imager that will allow it to perform both a spectroscopic and a photometric survey over $15\,000\,\deg^2$ of extra-Galactic sky \citep{Redbook}. The main aim of the mission is to measure the geometry of the Universe and the growth of structures up to redshift $z\sim 2$ and beyond. Euclid will include a photometric survey, measuring positions and shapes of over a billion galaxies, enabling the analysis of WL and GCph. 

\item SKA weak lensing survey is performed with the South-African mid-frequency array of the SKAO (see details in \cite{SKA:2018ckk}). The {\it continuum} survey identifies instead radio-emitting galaxies (e.g.\ star-forming or with an active radio galactic nucleus) from the reconstructed images from the interferometric data; their $z$ determination is poor, which make this survey the radio counterpart of photometric optical surveys. The reconstructed images of this survey are then used to obtain the weak lensing measurements. We report the specifications assumed for the continuum survey in \autoref{tab:ska-wl1}. 

\item The Vera C.\ Rubin Legacy Survey of Space and Time \cite{LSSTDarkEnergyScience:2018jkl} (hereafter Rubin), is a Stage IV galaxy survey using a ground-based telescope installed in Cerro Pachón in northern Chile. In this work we will consider Rubin for the photometric weak lensing and clustering probes, together with their cross-correlation, i.e. the 3$\times$2pt combination.
\end{itemize}

For each survey, we model the expected galaxy number density as 
\begin{equation}\label{eq:wl-nofz}
n(z) = \bar{n}_g \, z^2\,\exp\left[-\left(\frac{z}{z_0}\right)^\gamma\right] \;,
\end{equation}
where $\gamma$ and $z_0 = z_m/\sqrt{2}$ are given below for each survey in \autoref{tab:ska-wl1} where we list the specifications of the three adopted photometric surveys used for the forecasts in this work. \\

\begin{table}[htbp]
	\centering
	\begin{tabular}{ccccccc}
		 \multicolumn{7}{c}{Photo surveys} \\
		\hline
		\hline 
         & $f_{\rm sky}$ & $\bar n_g$  & $\epsilon_{\rm int}$ & $z_{m}$ & $\sigma_z^{\rm ph}/(1+z)$ & $\gamma$  \\
		\hline
         Euclid & $0.3636$ & $30$ & $0.30$  & $0.9$ & $0.05$  & $1.5$ \\         
         Rubin & $0.485$ & $27$ & $0.26$ & $0.156$ & $0.05$ & $0.68$ \\
         SKA II & $0.7272$ & $10$ & $0.30$ & $1.3$ & $0.05$ & $1.25$\\
		\hline 
	\end{tabular}
	\caption{ \label{tab:ska-wl1} Specifications for the WL surveys with Euclid \cite{}, Rubin \citep{LSST:2008ijt} and SKAO \citep{harrison_ska_2016}.
  }
\end{table}

\section{Forecasts for Euclid, Rubin and SKA}\label{results}

\begin{figure}
    \centering
           \includegraphics[width=0.45\linewidth]{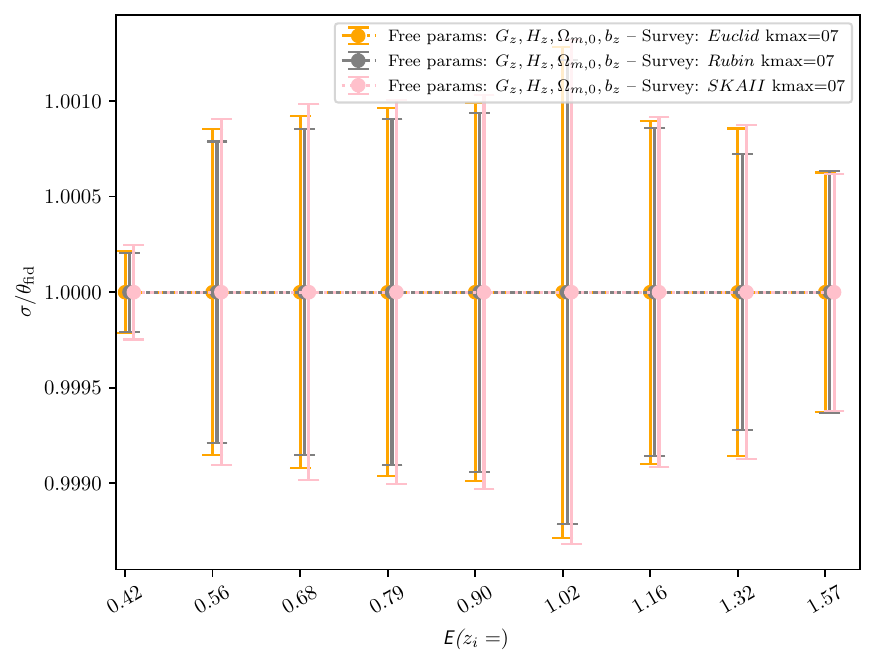}
           \includegraphics[width=0.45\linewidth]{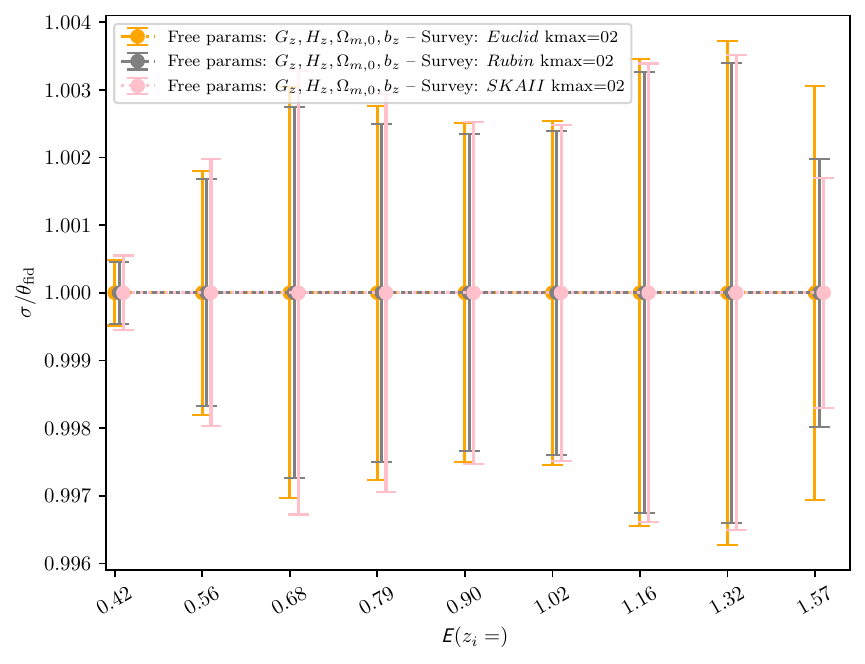}
           \includegraphics[width=0.45\linewidth]{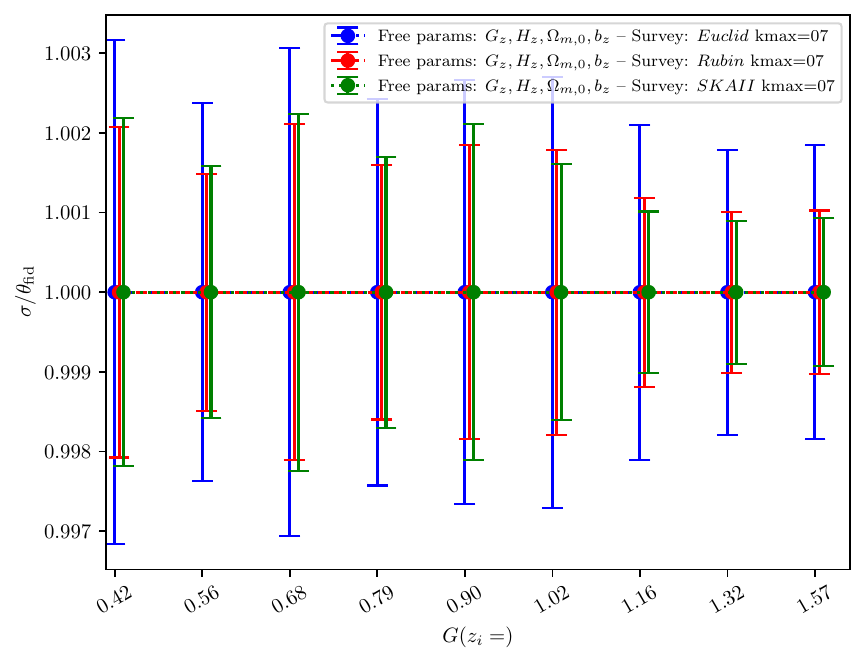}
           \includegraphics[width=0.45\linewidth]{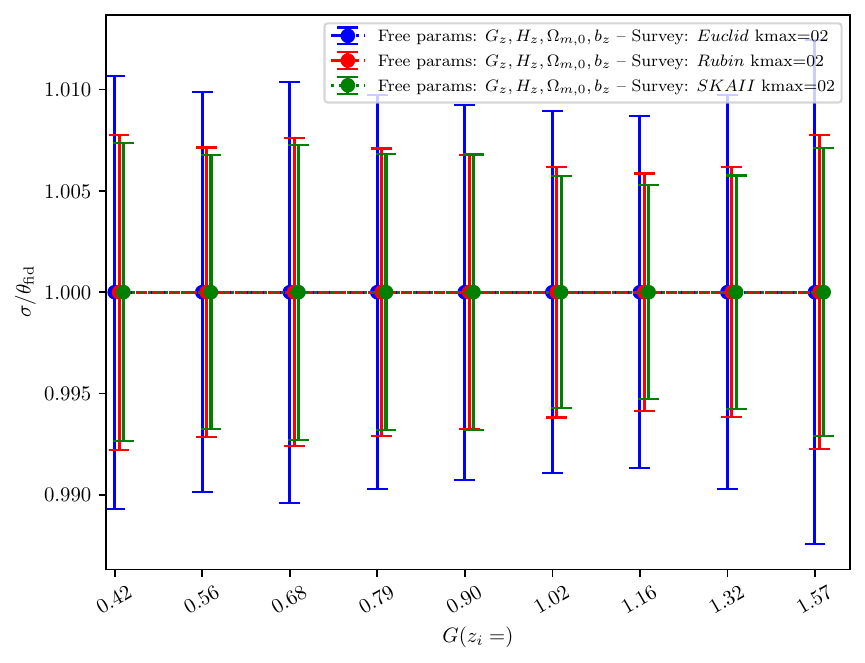}
            \includegraphics[width=0.45\linewidth]{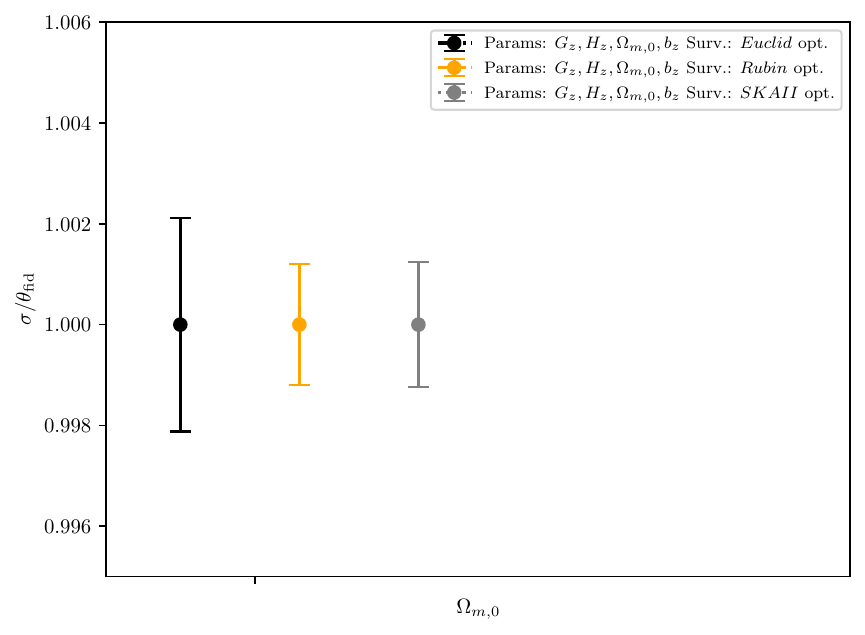}
           \includegraphics[width=0.45\linewidth]{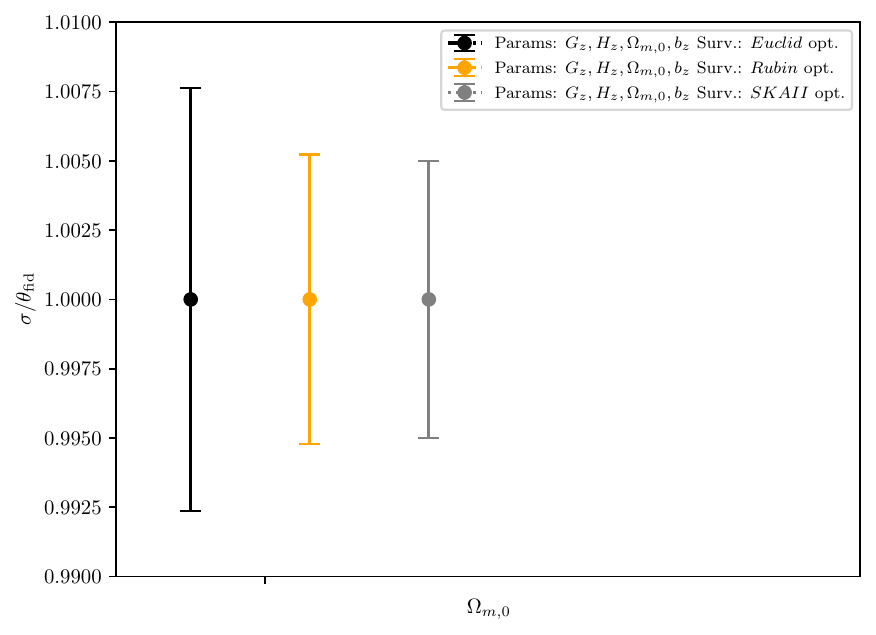} 
    \caption{1$\sigma$ marginalized forecasted errors on the expansion  $E(z_i)$, the growth $G(z_i)$ and the matter density $\Omega_{\rm m,0}$ parameters from Euclid, Rubin and SKA II photometric 3x2pt surveys when the galaxy biases are also left free to vary. Left panels are for when the optimistic settings are adopted, where the wavenumber values till $k_{max}=0.7$ are reached, for the right panels, pessimistic ones are adopted where the wavenumber values are limited to $k_{max}=0.2$ though still staying in the non linear regime.}
    \label{fig:GzEzOm0bz}
\end{figure}

We start by showing in Fig.~\ref{fig:GzEzOm0bz} the constraints obtained on the expansion and the growth parameters in the first case where we only vary, in the most constraining scheme of this work, the matter density, the expansion, the growth and the tracers bias parameters, and that in the optimistic (left panel) where $k_{max}=0.7$, and pessimistic (right panel) settings where $k_{max}=0.2$, with forecasts for the three adopted photometric surveys, Euclid, Rubin and SKA. The errors are within the sub-percent for the optimistic, with constraints from further including non-linear scales, and degrade by one order of magnitude to the percent in the pessimistic case for all the redshift bins and regardless of the surveys with a small advantage for Rubin over the remaining surveys despite its limitation here in terms of low redshift windows, which is due to its higher sky coverage while the galaxy density and lensing error are still close to the Euclid survey. 
We notice also that the expansion and the growth parameters are less constrained at intermediate than high redshifts for the pessimistic or the optimistic case while the lower redshifts are the most constraining. This could be understood for the expansion as the translation of the interplay of the density of the source and lensed galaxies through the lensing efficiency while the growth could be moreover influenced by its impact on the power spectrum with growth reverting to GR usually at high redshift lowering the impact of our MG parameters.

\begin{figure}
    \centering
           \includegraphics[width=0.45\linewidth]{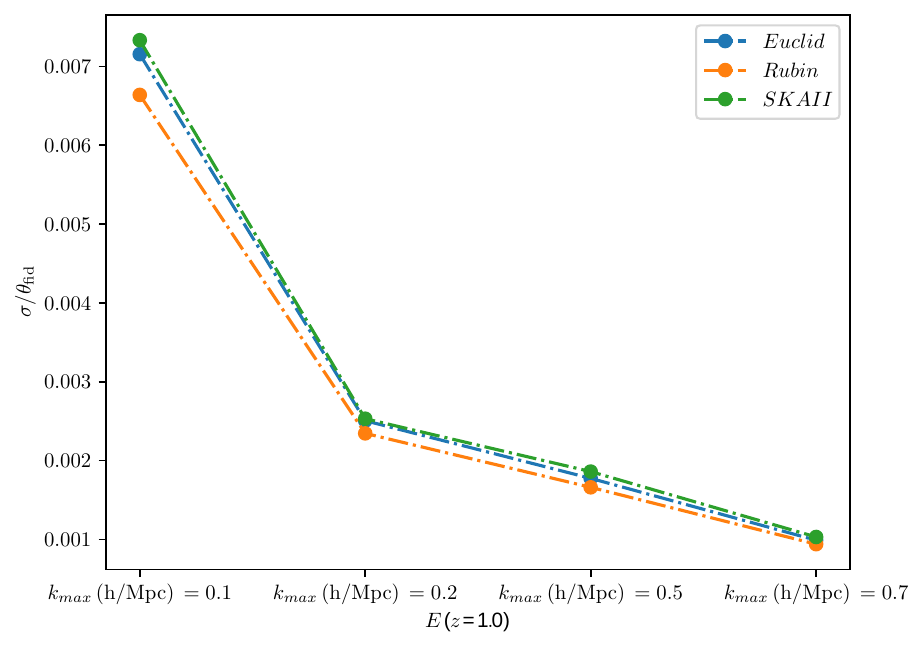}
           \includegraphics[width=0.45\linewidth]{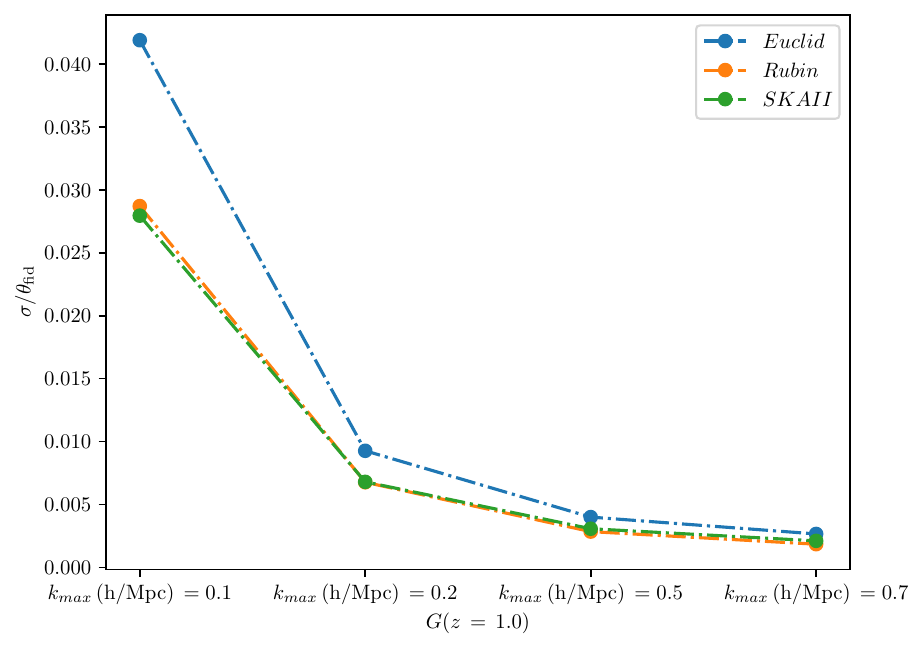}
           
    \caption{Showing the variation of the 1$\sigma$ marginalized forecasted relative errors for different wavenumbers for the growth and expansion factors for the same cases we considered when producing Fig.~\ref{fig:GzEzOm0bz} }
    \label{fig:1}
\end{figure}

\begin{figure}
    \centering
    \includegraphics[width=0.45\linewidth]{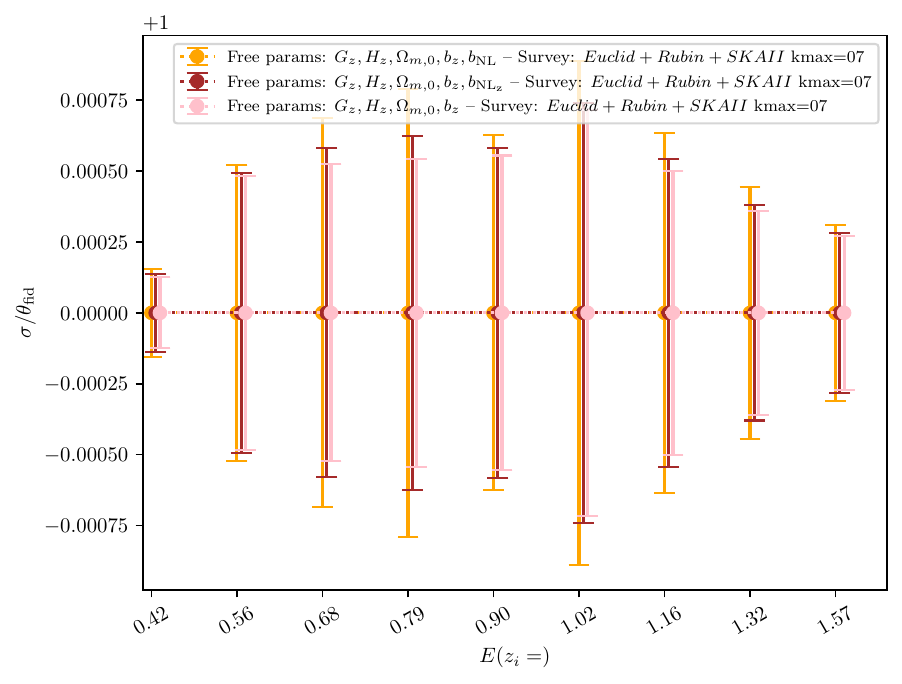}
    \includegraphics[width=0.45\linewidth]{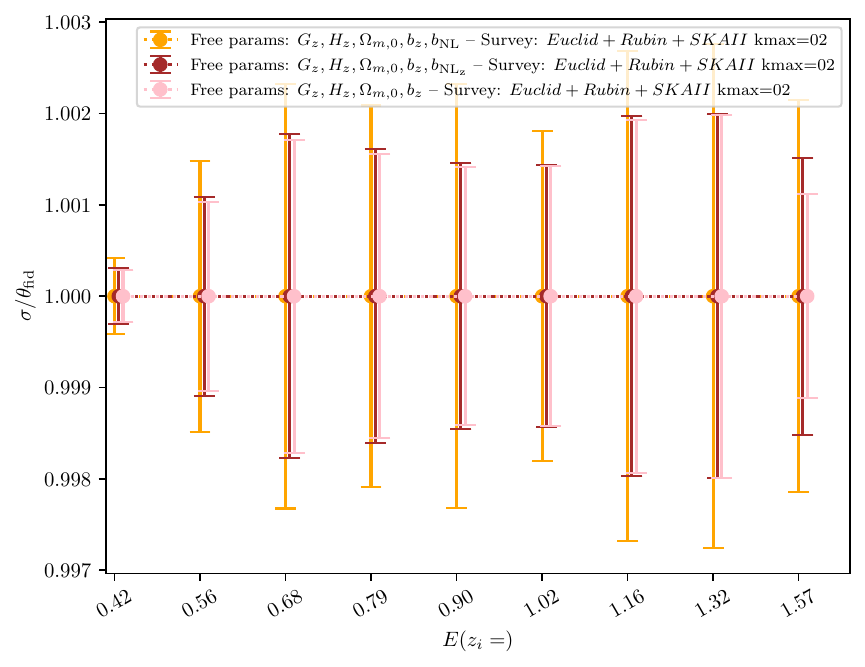}
    \includegraphics[width=0.45\linewidth]{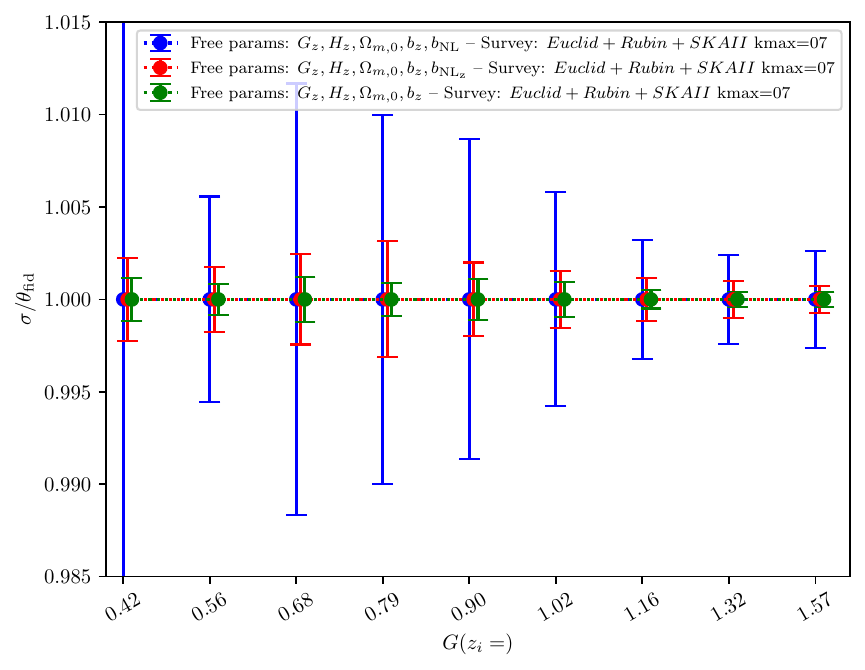}
    \includegraphics[width=0.45\linewidth]{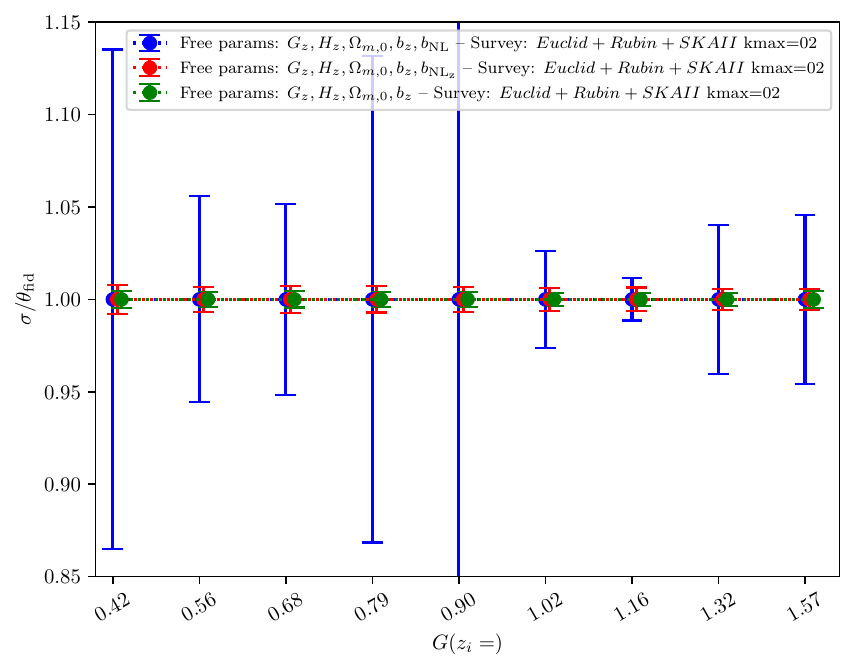}
    \includegraphics[width=0.45\linewidth]{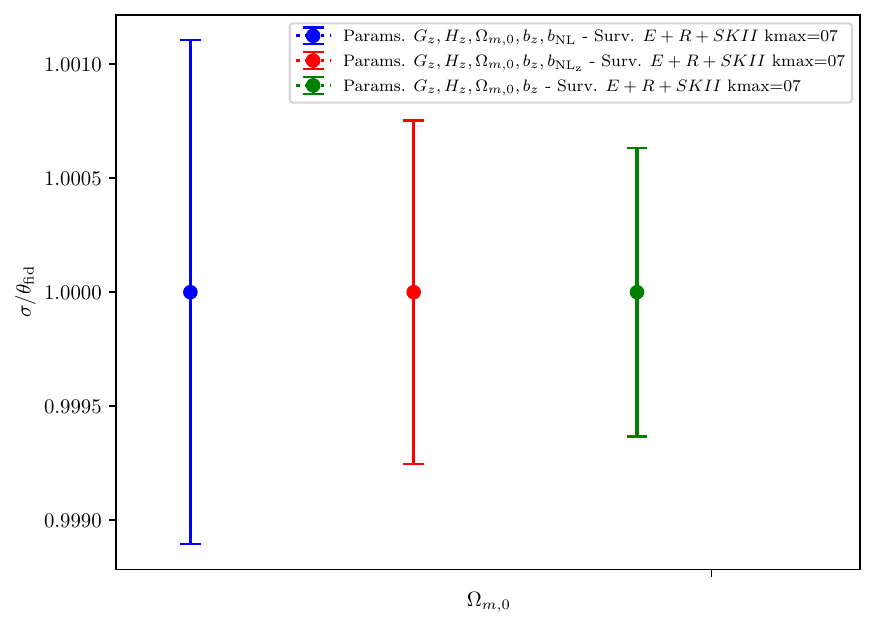}
    \includegraphics[width=0.45\linewidth]{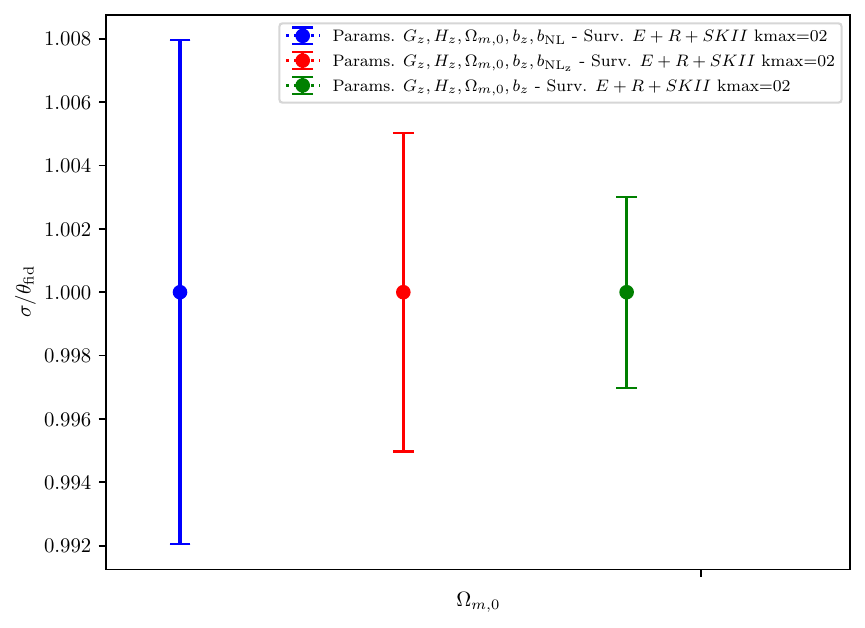}
    \caption{1$\sigma$ marginalized forecasted errors on the expansion $E(z_i)$, the growth $G(z_i)$ and the matter density $\Omega_{\rm m,0}$ parameters from the combination of Euclid, Rubin and SKA II photometric 3x2pt surveys when the galaxy biases are left free to vary in comparison to two other cases where, in the first, the non linear halo bias parameters $B_{\rm NL}$ are left free supposing a redshift dependence, and in the second, a further halo mass dependence was considered for $B_{\rm NL}$. Left panels are for when the optimistic settings are adopted, where the wavenumber values till $k_{max}=0.7$ are reached, for the right panels, pessimistic ones are adopted where the wavenumber values are limited to $k_{max}=0.2$ though still staying in the non linear regime.}
    \label{fig:GzEzOm0bzBNLz}
\end{figure}

Then we show in Fig~\ref{fig:GzEzOm0bzBNLz} constraints on the same previously varied parameters but by further allowing the non halo bias to vary. This is expecting to lower our constraints in general, so we try to remedy by showing constraints from combining the three surveys at once. We keep for comparison the case where $b_{\rm NL}$ was fixed similar to the previous case and add to it a case where $b_{\rm NL}$ is parameterised as z dependent following the same division of $E(z_i)$ and $G(z_i)$ and a more free case where $b_{\rm NL}$ is additionally parameterised as function of the halo mass, which should open further the parameter space and is expected to further weaken our constraints. As expected we observe first that our previous basic case is showing stronger constraints due to the fact that we are showing constraints from combining the three surveys. Then we observe a weakening of the constraints resulting from the more freedom we gave to BNL. We notice that in the case of $b_{\rm NL}$ with mass and z dependence that the $E(z_i)$ constraints are showing the same pattern as before since their impact is supposed to not be dependent from $b_{\rm NL}$ with however a little decrease with high z since $E(z_i)$ also enters the equation of growth. While the stronger impact will of course be seen on the growth parameters. The same trend is observed for the matter density that directly impacts the non linear modelling in the one halo equations, or the $b_{\rm NL}$ with relative high impact at small scales reached by the optimistic case.

\begin{figure}
    \centering
           \includegraphics[width=0.45\linewidth]{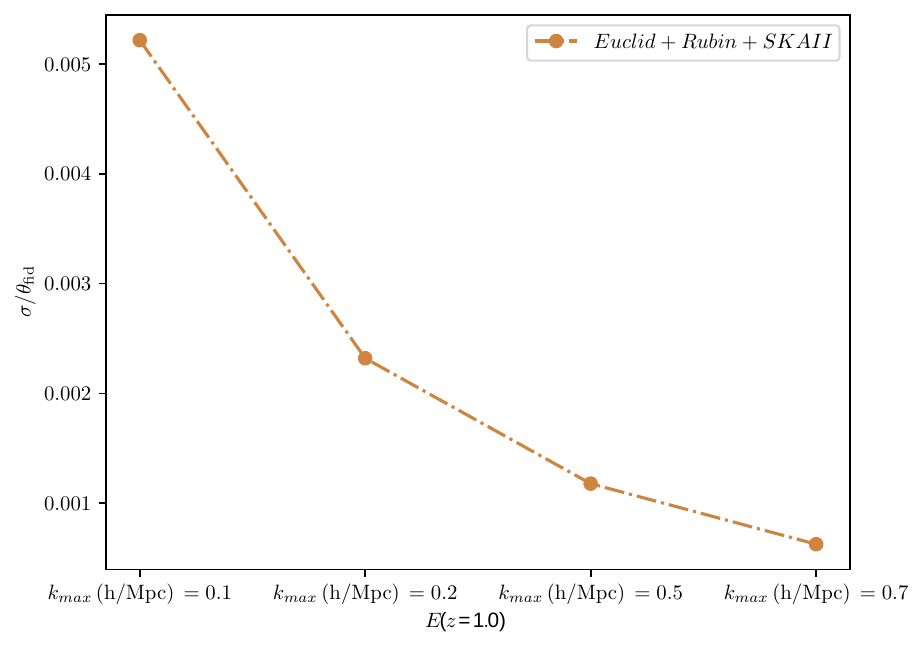}
           \includegraphics[width=0.45\linewidth]{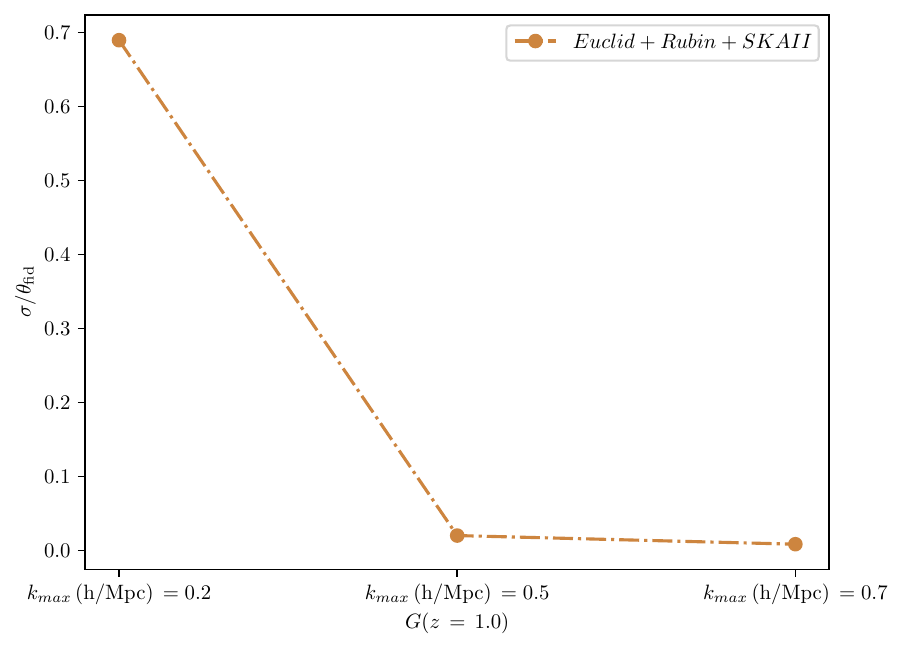}

           \includegraphics[width=0.45\linewidth]{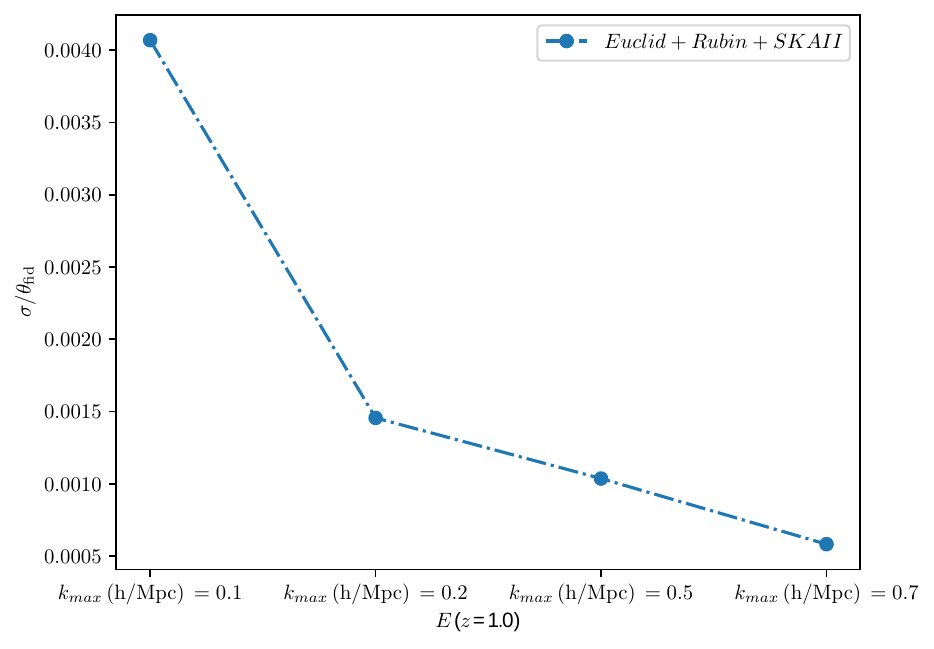}
           \includegraphics[width=0.45\linewidth]{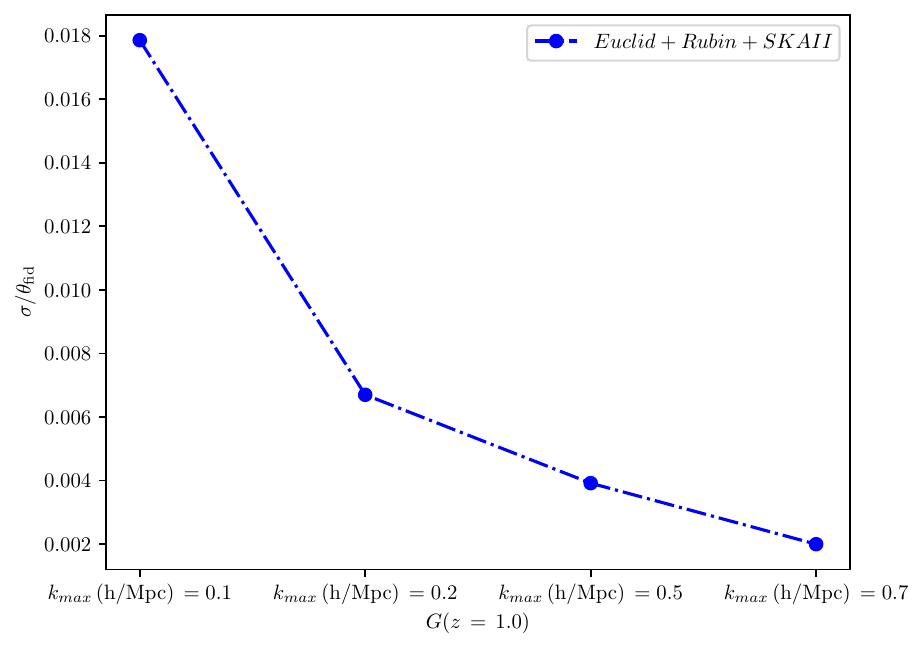}

    \caption{Showing the variation of the 1$\sigma$ marginalized forecasted relative errors for different wavenumbers for the growth and expansion factors for the same cases we considered when producing Fig.~\ref{fig:GzEzOm0bzBNLz} }
    \label{fig:2}
\end{figure}

\begin{figure}
    \thisfloatpagestyle{empty}
    \centering
    \includegraphics[width=0.45\linewidth]{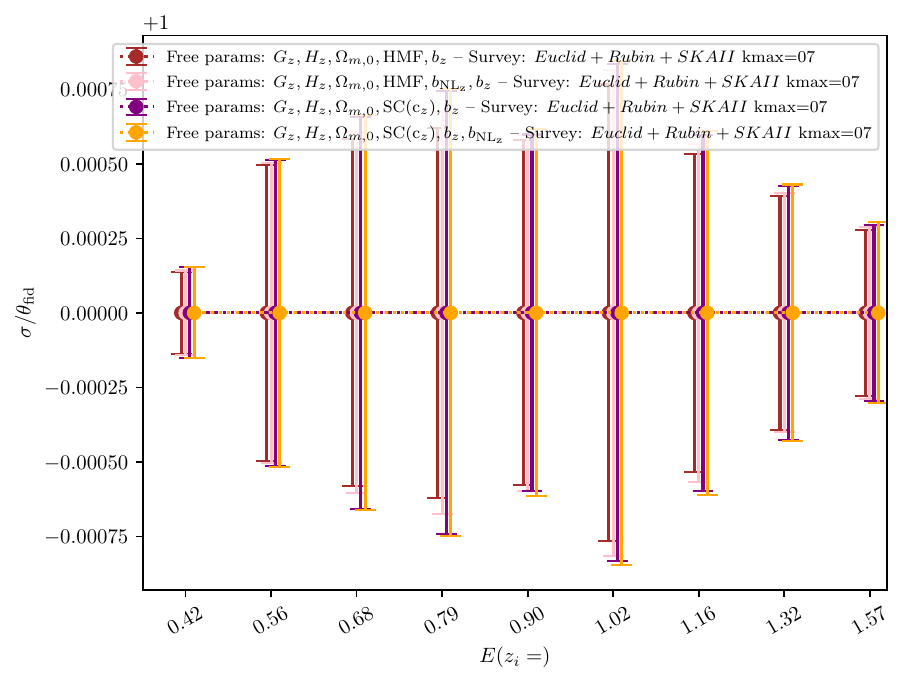}
    \includegraphics[width=0.45\linewidth]{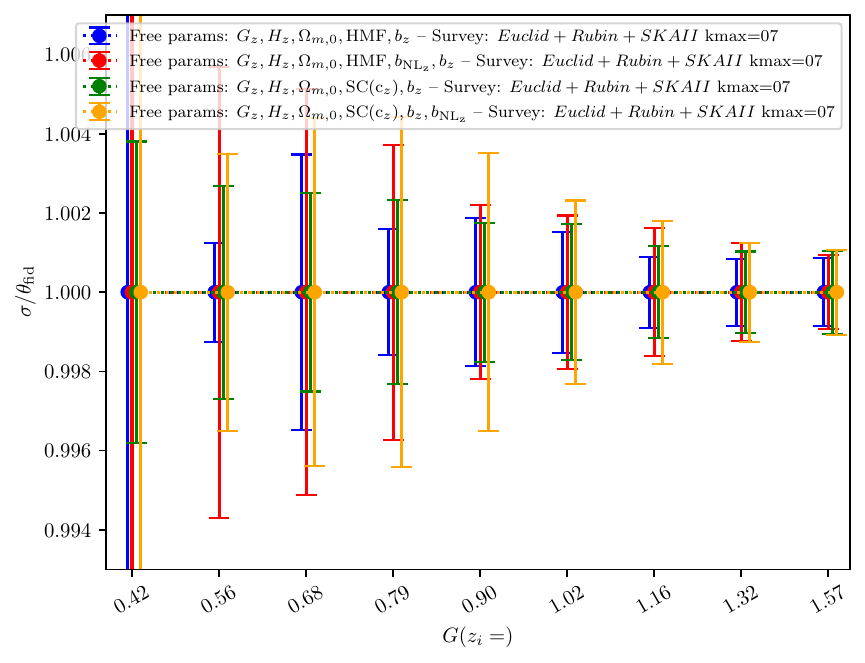}
    \includegraphics[width=0.45\linewidth]{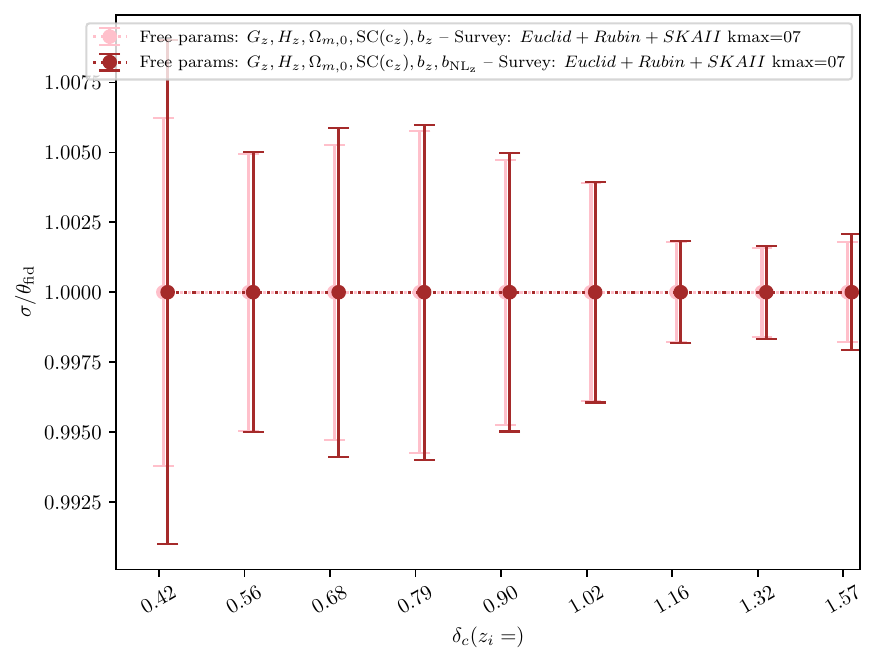}
    \includegraphics[width=0.45\linewidth]{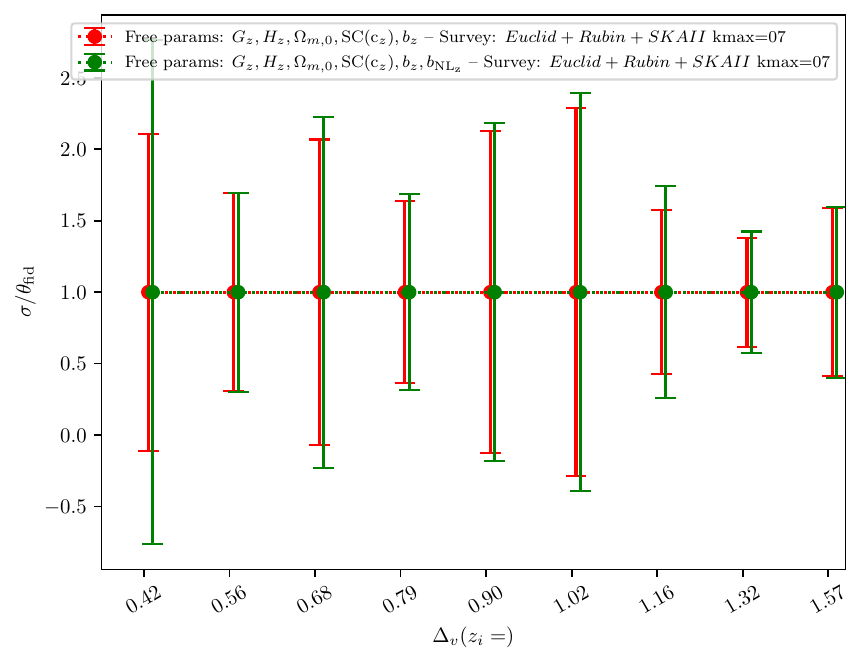}
    \includegraphics[width=0.45\linewidth]{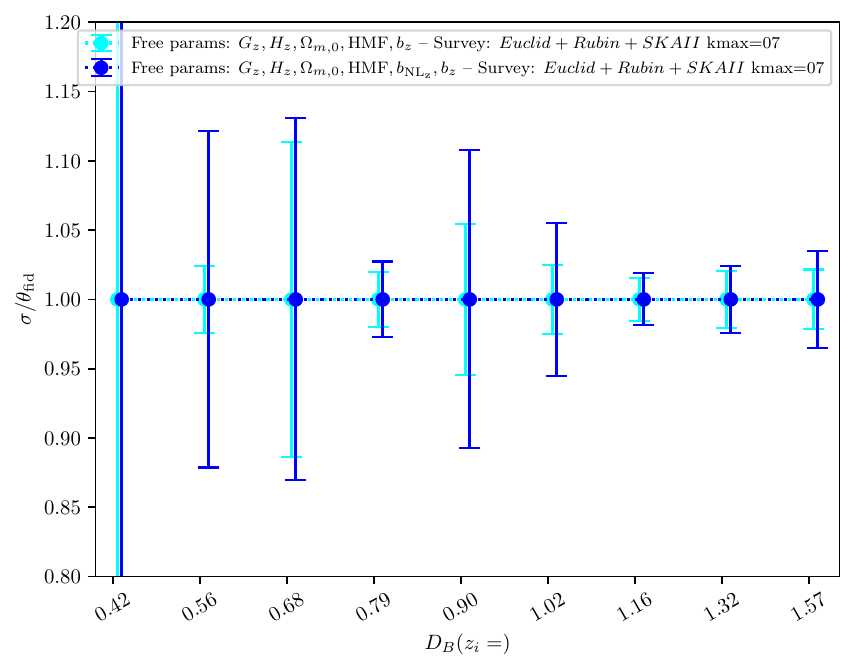}
    \includegraphics[width=0.45\linewidth]{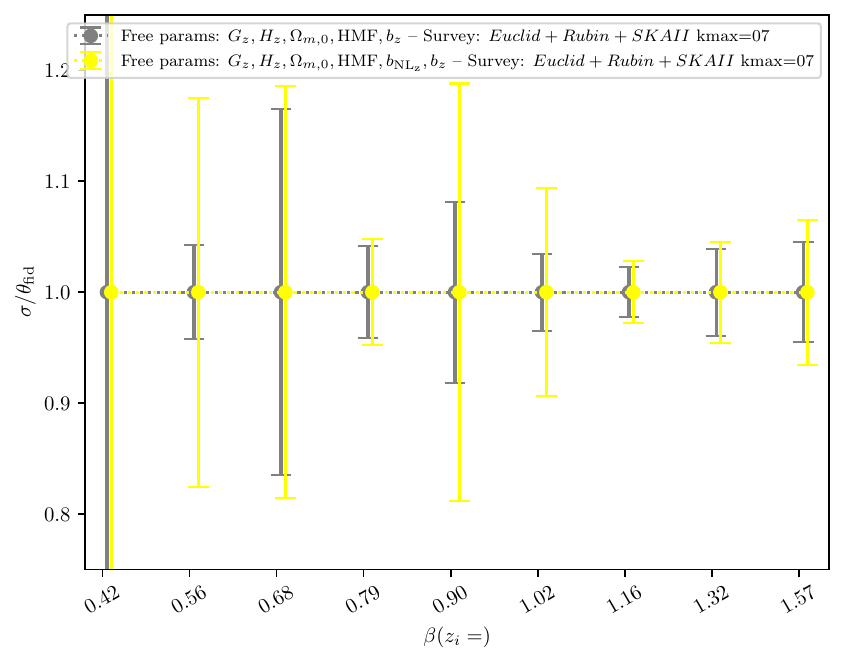}
    \includegraphics[width=0.45\linewidth]{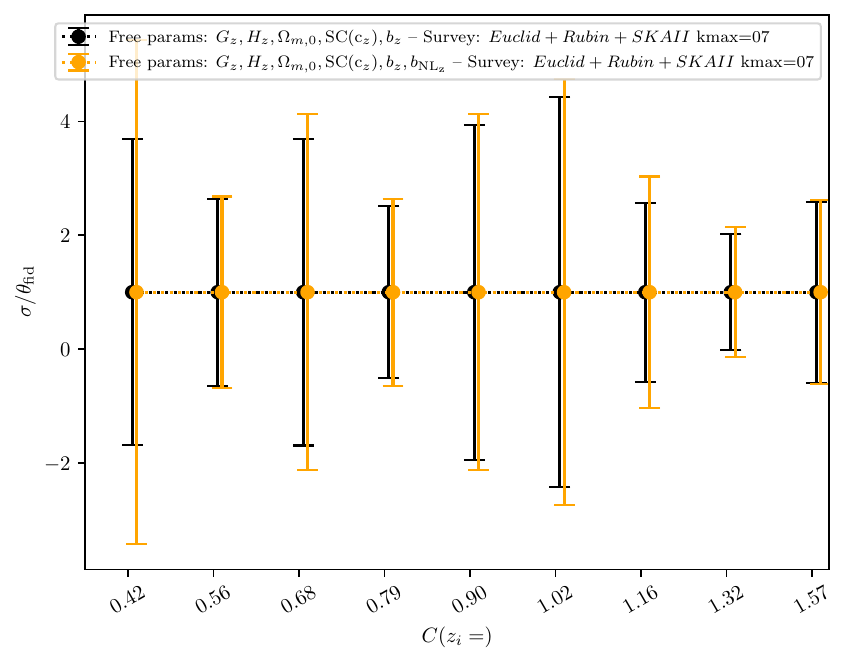}
    \includegraphics[width=0.45\linewidth]{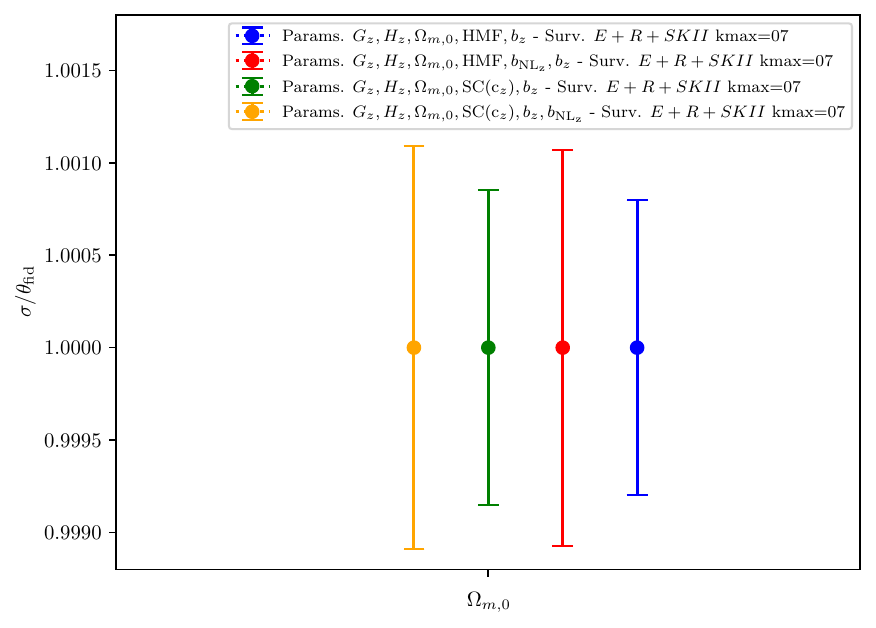}    
    \caption{1$\sigma$ marginalized forecasted errors on the expansion $E(z_i)$, the growth $G(z_i)$ and the matter density $\Omega_{\rm m,0}$ parameters from the combination of Euclid, Rubin and SKA II photometric 3x2pt surveys when the galaxy biases  are left free to vary, in comparison to a case where the halo model excursion set parameters are also free. In each of this two schemes, we also considered two cases, where in the first, the non linear halo bias parameters $B_{\rm NL}$ are left free supposing a redshift dependence, and in the second, a further halo mass dependence was considered. All panels are for when the optimistic settings are adopted, where the wavenumber values till $k_{max}=0.7$ are reached.}    \label{fig:GzEzOm0HMFbNLzbz__GzEzOm0SCSzbz}
\end{figure}

\begin{figure}
    \centering
           \includegraphics[width=0.45\linewidth]{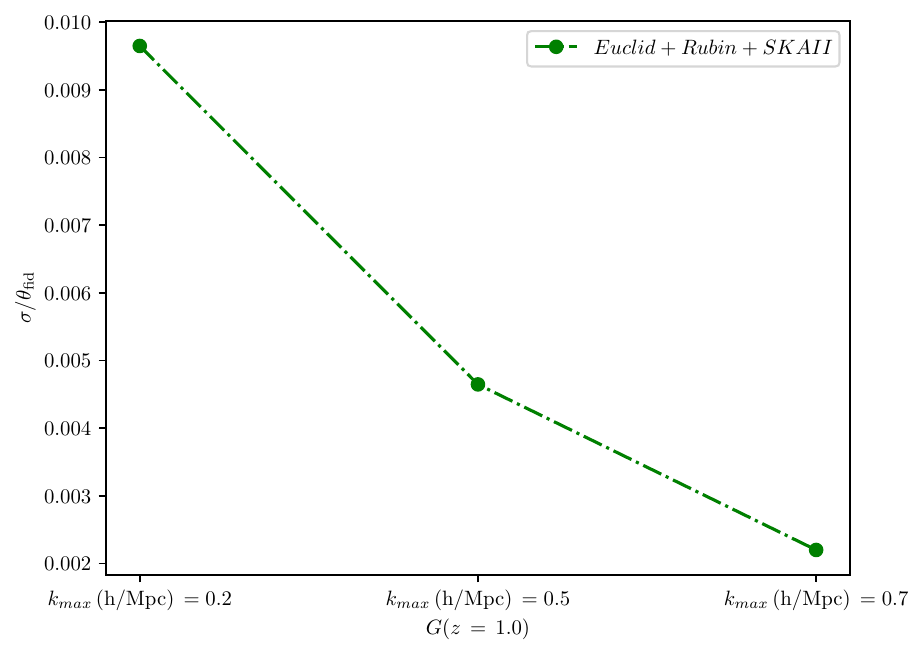}
           \includegraphics[width=0.45\linewidth]{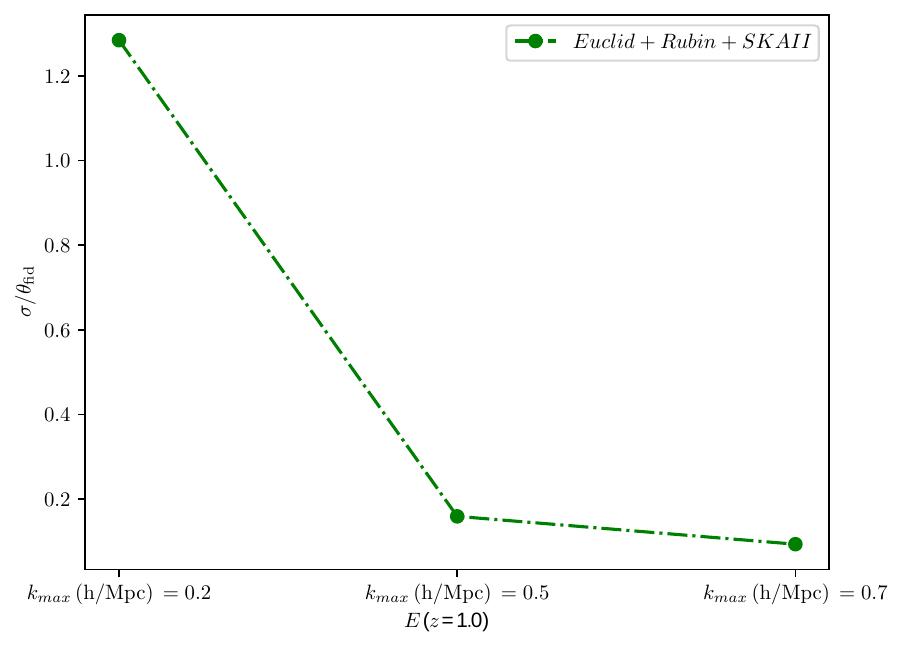}

           \includegraphics[width=0.45\linewidth]{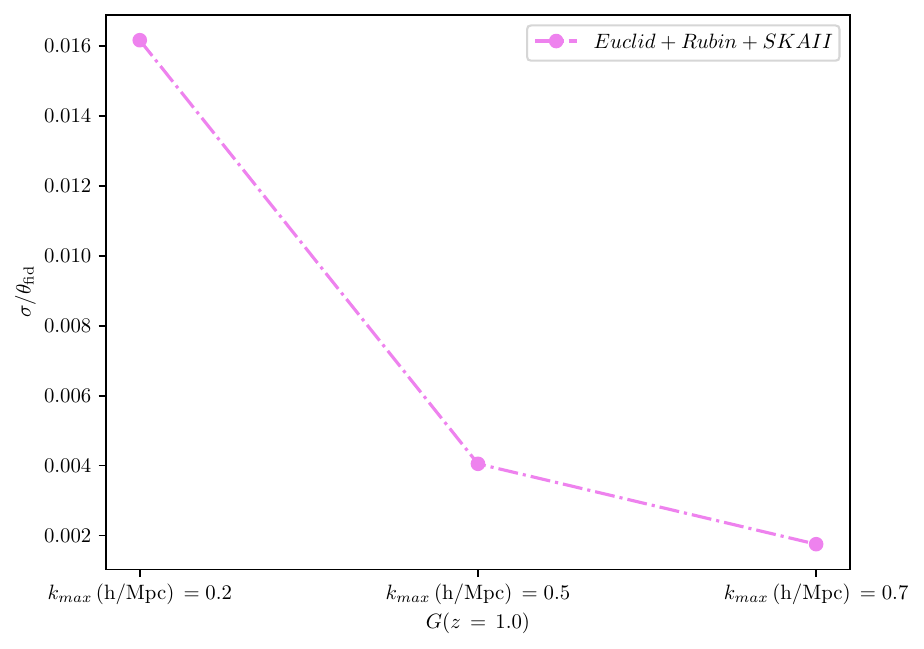}
           \includegraphics[width=0.45\linewidth]{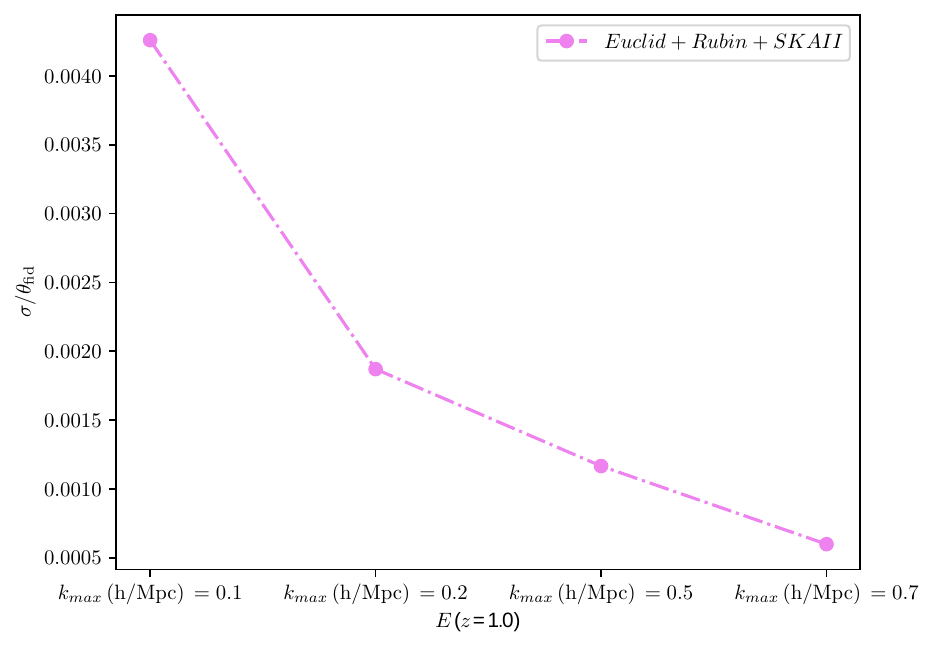}

           \includegraphics[width=0.45\linewidth]{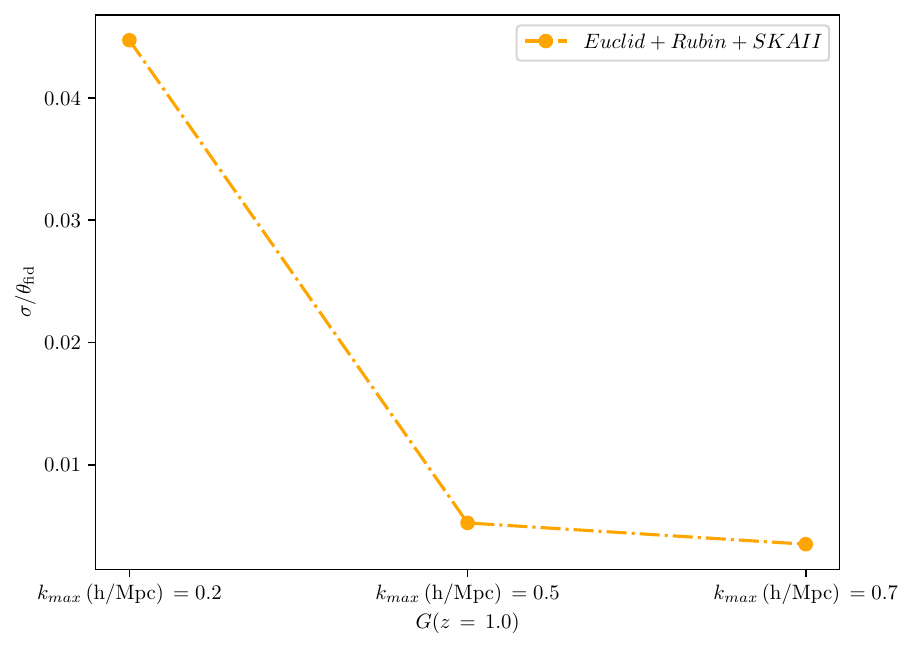}
           \includegraphics[width=0.45\linewidth]{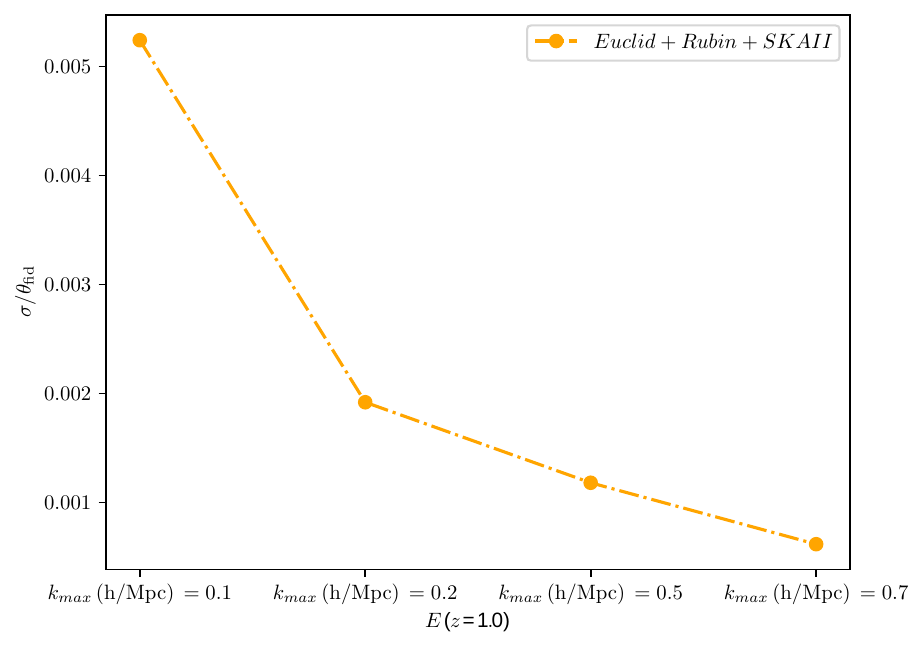}
                      
    \caption{Showing the variation of the 1$\sigma$ marginalized forecasted relative errors for different wavenumbers for the growth and expansion factors for the same cases we considered when producing Fig.~\ref{fig:GzEzOm0HMFbNLzbz__GzEzOm0SCSzbz}}
    \label{fig:3}
\end{figure}

To asses next the impact of the halo mass function parameters along with the spherical collapse parameters on the expansion and growth ones, we left the former parameters as free to vary and consider a high optimistic case, in the hope to break degeneracies from the many degree of freedom we introduced. We show in Fig.~\ref{fig:GzEzOm0HMFbNLzbz__GzEzOm0SCSzbz} constraints on all our paramters in the case where we suppose we reach the wavenumber values till $k_{max}=0.7$, but we also show in Fig.~\ref{fig:3} the constraints on the main growth and expansion parameters for three less constraining higher scales. In more details, we show in Fig.~\ref{fig:GzEzOm0HMFbNLzbz__GzEzOm0SCSzbz} marginalized forecasted errors on the expansion $E(z_i)$, the growth $G(z_i)$ and the matter density $\Omega_{\rm m,0}$ parameters from the combination of Euclid, Rubin and SKA II photometric 3x2pt surveys when the galaxy biases are left free to vary, in comparison to a case where the halo model excursion set parameters are also free. In each of this two schemes, we also considered two cases, where in the first, the non linear halo bias parameters $B_{\rm NL}$ are left free supposing a redshift dependence, and in the second, a further halo mass dependence was considered.  

\clearpage{}

\begin{figure}
    \centering
    \includegraphics[width=0.45\linewidth]{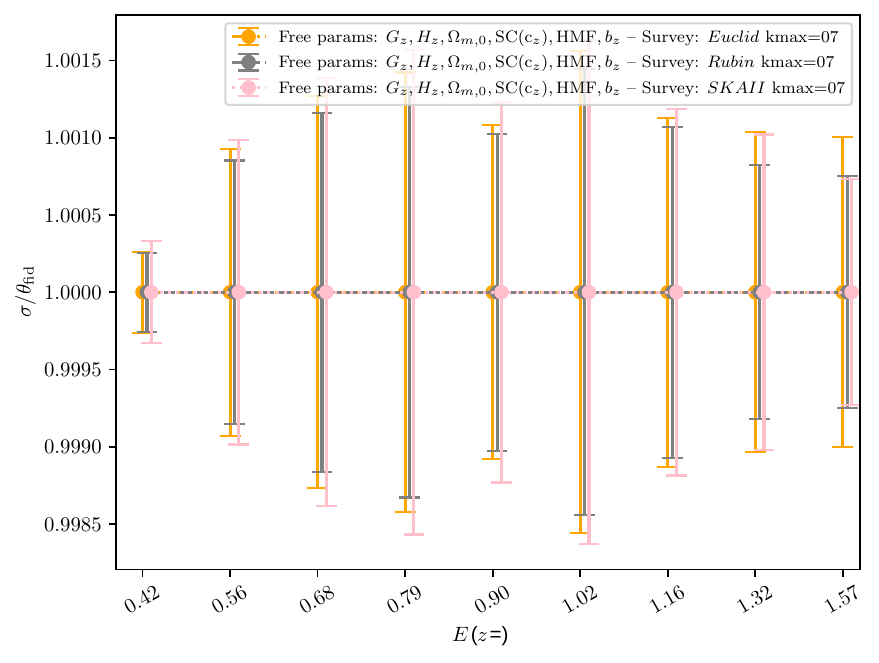}
    \includegraphics[width=0.45\linewidth]{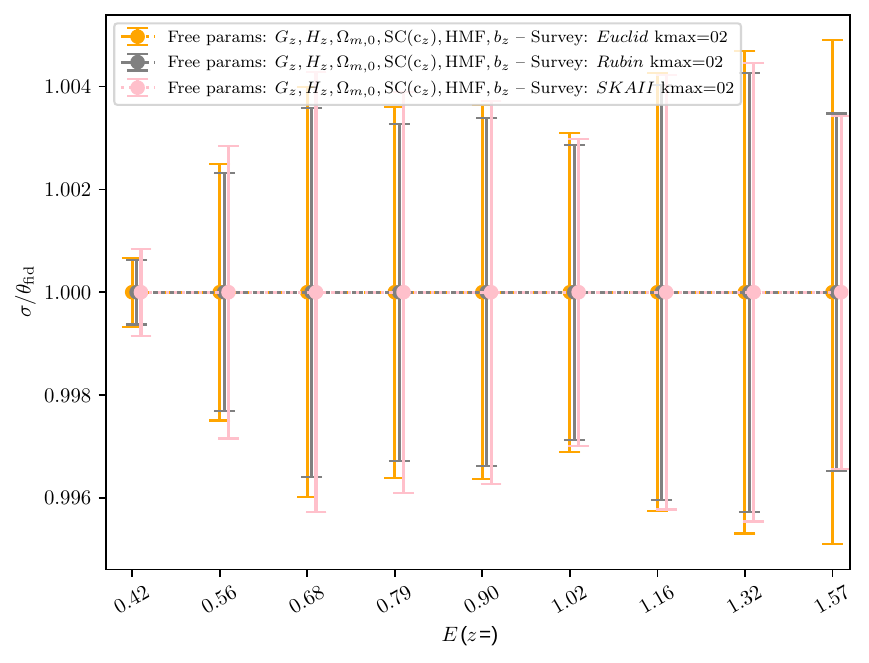}
    \includegraphics[width=0.45\linewidth]{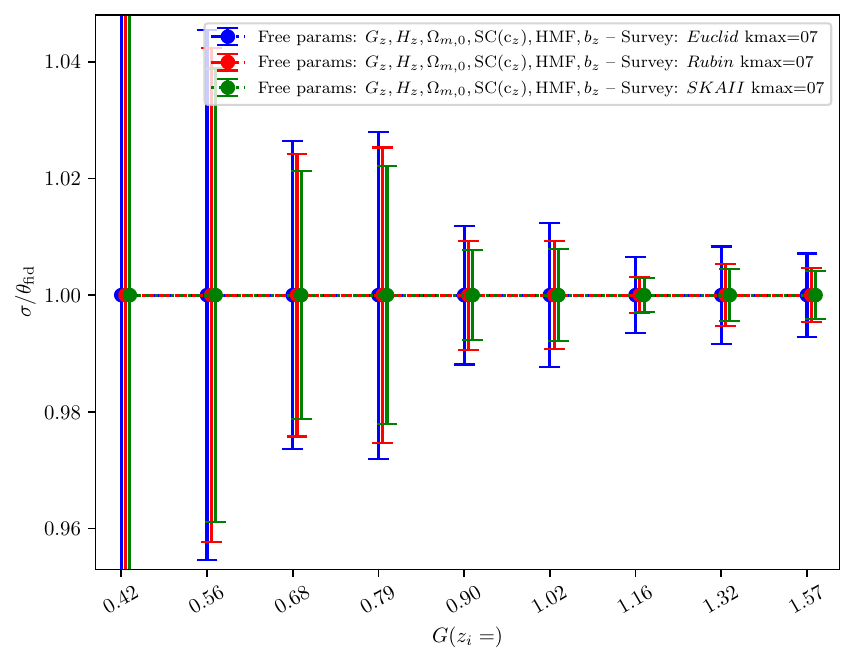}
    \includegraphics[width=0.45\linewidth]{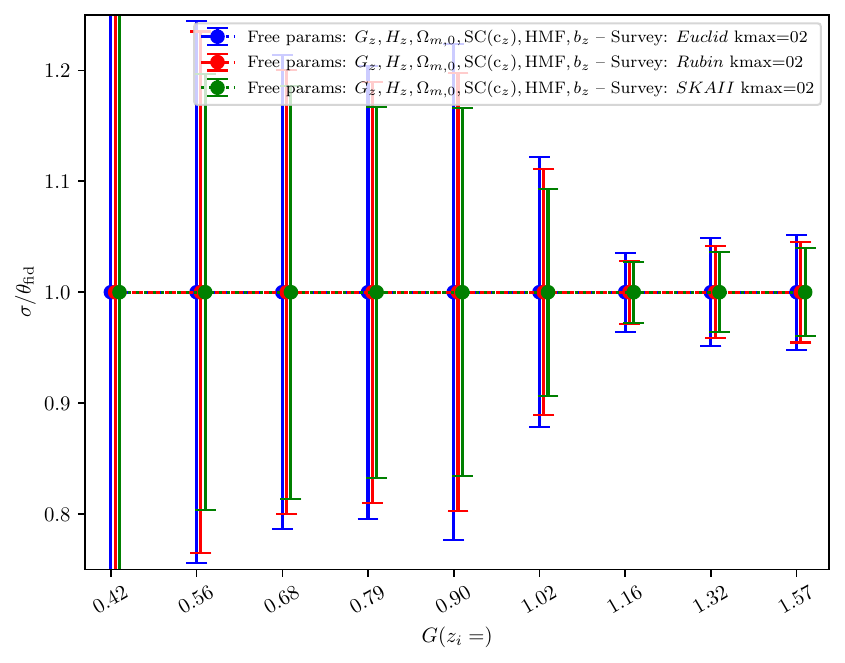}
     \includegraphics[width=0.45\linewidth]{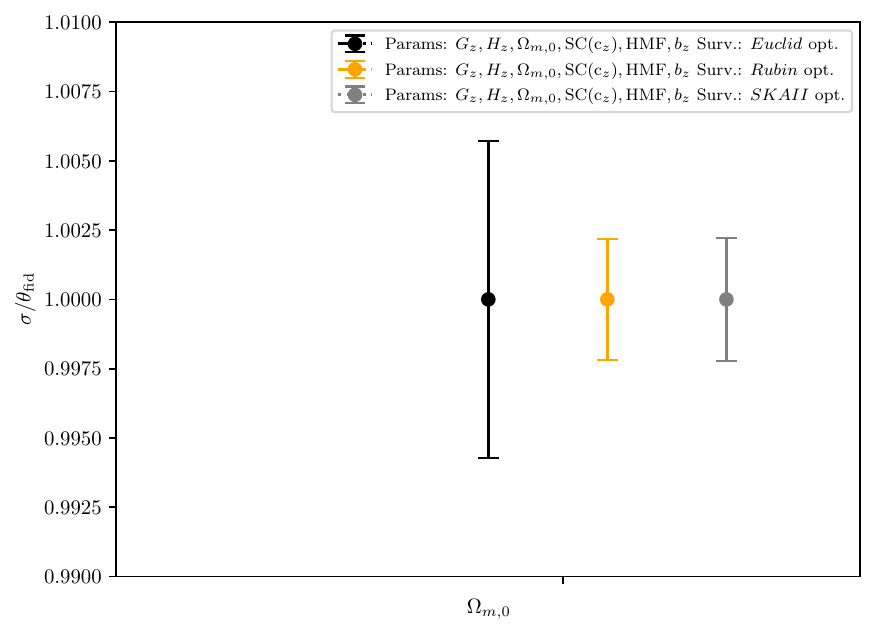}
    \includegraphics[width=0.45\linewidth]{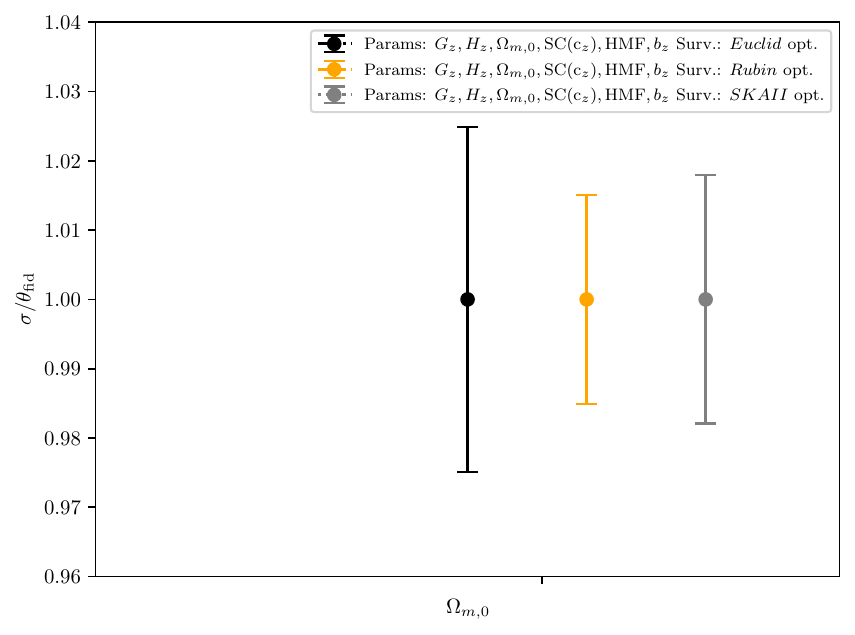}
    \caption{1$\sigma$ marginalized forecasted errors on the expansion  $E(z_i)$, the growth $G(z_i)$ and the matter density $\Omega_{\rm m,0}$ parameters from each of the photometric 3x2pt surveys: Euclid, Rubin or SKA II,  when the galaxy biases are left free to vary along with the non linear halo bias parameters $B_{\rm NL}$ while supposing a redshift dependence for the latter, along with the excursion set halo mass function parameters with redhsift dependence for the concentration parameters, and the spherical collapse model parameters. Left panels are for when the optimistic settings are adopted, where the wavenumber values till $k_{max}=0.7$ are reached, for the right panels, pessimistic ones are adopted where the wavenumber values are limited to $k_{max}=0.2$ though still staying in the non linear regime.}
    \label{fig:GzEzOm0SCSzHMFbz}
\end{figure}

\begin{figure}
    \centering
           \includegraphics[width=0.45\linewidth]{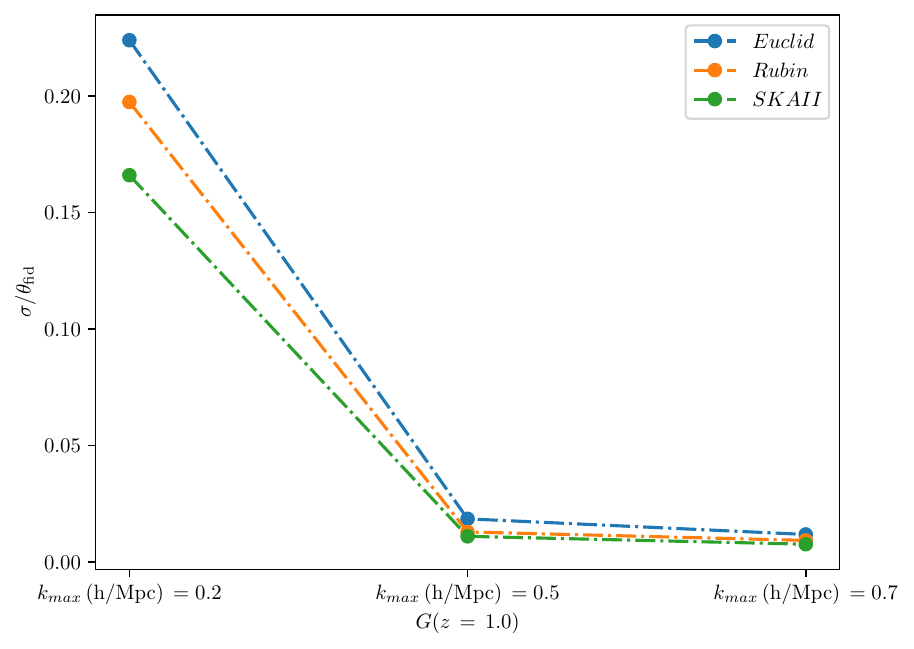}
           \includegraphics[width=0.45\linewidth]{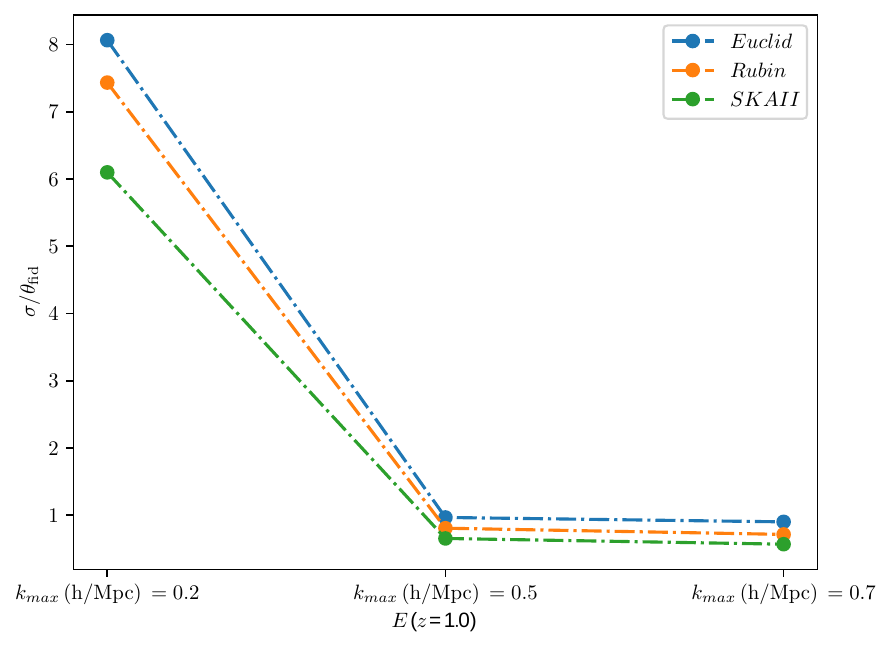}
                      
    \caption{Showing the variation of the 1$\sigma$ marginalized forecasted relative errors for different wavenumbers for the growth and expansion factors for the same cases we considered when producing Fig.~\ref{fig:GzEzOm0SCSzHMFbz}}
    \label{fig:4}
\end{figure}

\begin{figure}
    \centering
    \includegraphics[width=0.45\linewidth]{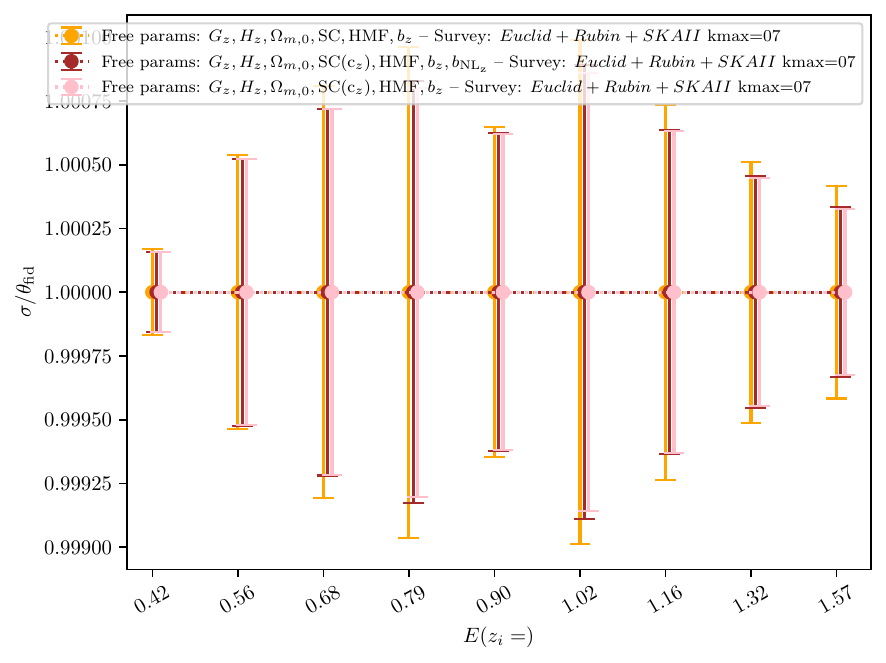}
    \includegraphics[width=0.45\linewidth]{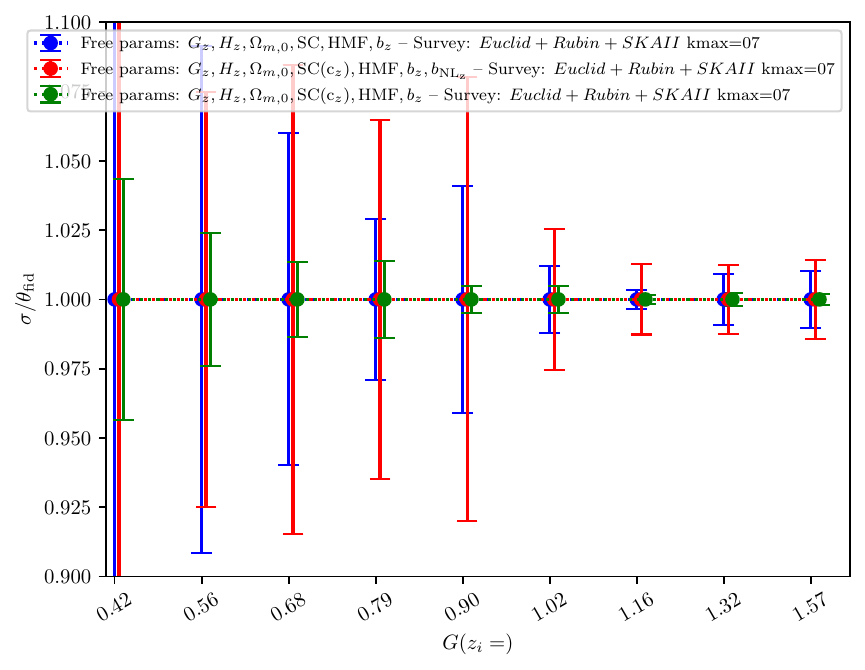}
    \includegraphics[width=0.45\linewidth]{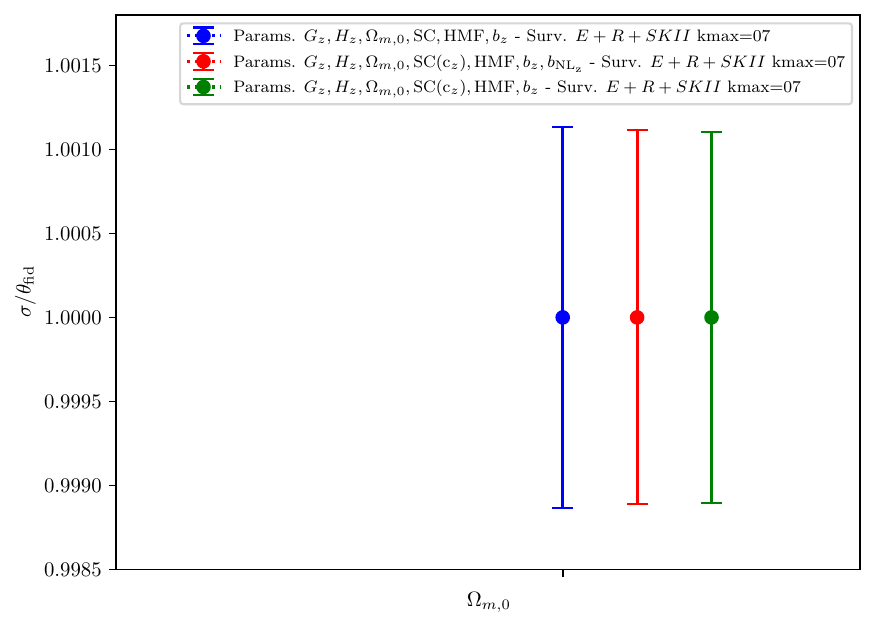}
    \caption{1$\sigma$ marginalized forecasted errors on the expansion $E(z_i)$, the growth $G(z_i)$ and the matter density $\Omega_{\rm m,0}$ parameters from the combination of Euclid, Rubin and SKA II photometric 3x2pt surveys when the galaxy biases are left free to vary in all cases along with, in the first case, further leaving the spherical collapse and the halo model, including the concentration parameters with redshift dependence, as free while the non linear bias are fixed to their fiducials; in the second case the concentration parameters were parameterized as function of redshift and mass; and in the third case the non linear halo bias parameters were additionally left free  while keeping the redshift dependence for the concentration parameter. All panels are for when the optimistic settings are adopted, where the wavenumber values till $k_{max}=0.7$ are reached.}
    \label{fig:GzEzOm0SCSzHMFbzBNLz}
\end{figure}

\begin{figure}
    \centering
           \includegraphics[width=0.45\linewidth]{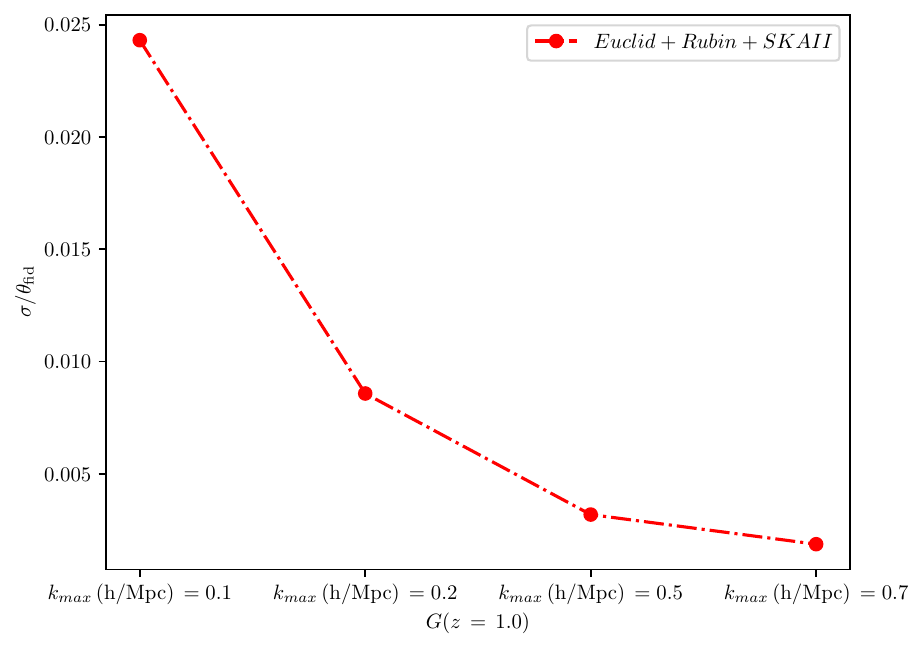}
           \includegraphics[width=0.45\linewidth]{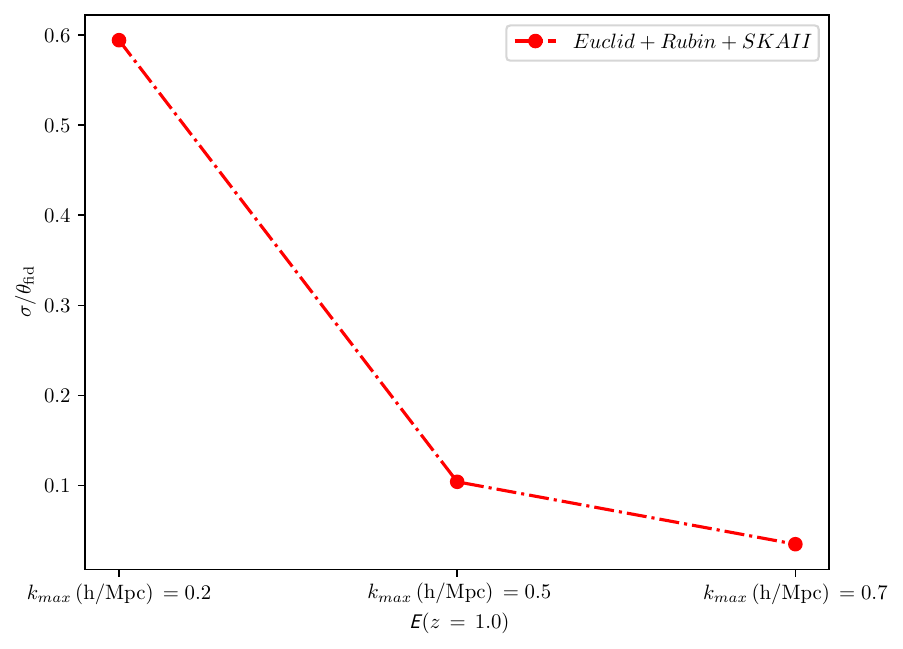}
                      
    \caption{Showing the variation of the 1$\sigma$ marginalized forecasted relative errors for different wavenumbers for the growth and expansion factors for the same cases we considered when producing Fig.~\ref{fig:GzEzOm0SCSzHMFbzBNLz}}
    \label{fig:4}
\end{figure}

\section{Conclusions}\label{conclusion}

\hspace{15pt} Present and future large-scale cosmological data are opening a new era of precision while at the same time showing how difficult is to improve upon accuracy, given not just the observational systematics but also the many uncertainties about theoretical modelling. 

In this work, we introduced a methodology to extend the 3$\times$2pt galaxy clustering and lensing probes Fisher matrix forecasts to non-linear scales without the need of selecting a cosmological model, following a complementary route, where instead of focusing on specific cosmological models, we develop a methodology to extract from the data information that is as much model-independent as possible. More specifically, we do not assume a cosmological model neither at the background nor at the perturbed level, and parametrize galaxy/matter biasing on general grounds, via the parameterised bias expansion. For that, we make use of standard halo model and excursion set theory and, instead of choosing a specific model, we parametrize the linear power spectrum and the growth rate in several $k$ and $z$ bins. We use a Fisher matrix analysis extended, in which the parameters are not the usual cosmological ones, but rather the linear growth factor $G(z)$ of the linear power spectrum and the background expansion factor $E(z)$ in wavenumber and redshift bins, plus free functions of linear and non linear halo bias, the halo model principal ingredients, i.e. the ones related to the spherical collapse or the halo mass function modelling. We show that one can then obtain model-independent constraints of the expansion $E(z)=H(z)/H_0$ and the growth factor $G(z)$, besides to the different bias and non-linear modelling functions. We apply the technique to Euclid, Rubin and SKA public specifications in the range $0.2\le z \le 1.8$ and show the change in gain in precision at each $z$-shell when going from optimistic high linear scales to more pessimistic settings, or between  each survey  or a combination of them all. In the most agnostic case one can still reach high relative precision on $E(z)$ in the order of the percent level in the optimistic and combining the three surveys at once while the growth factor has for the same settings one order of magnitude weaker constraints. We also showed how neglecting the non-linear corrections can have a large effect on the constraints.

\section*{Acknowledgments}

ZS acknowledges support from DFG project 456622116 at the time when this study was conducted and support from the research projects PID2021-123012NB-C43, PID2024-159420NB-C43, the Proyecto de Investigación SAFE25003 from the Consejo Superior de Investigaciones Científicas (CSIC), and the Spanish Research Agency (Agencia Estatal de Investigaci\'on) through the Grant IFT Centro de Excelencia Severo Ochoa No CEX2020-001007-S, funded by MCIN/AEI/10.13039/501100011033. 

\appendix

\bibliographystyle{JHEP2015}

\bibliography{references,scaling_bib}
\label{lastpage}
\end{document}